%% file: popl254-elango.tex
\documentclass{sigplanconf}
\usepackage{pslatex} % Better than Times (codes are awful...)
\usepackage{setspace}
\usepackage{multirow}

\usepackage[linesnumbered]{algorithm2e}
\usepackage{amsmath,amssymb}
\usepackage{wrapfig}
\usepackage{subfigure}
\usepackage{subfig}
\usepackage{paralist}
\usepackage{alltt,url}
\usepackage{xspace}
\usepackage{epsfig}
\usepackage{pstricks,graphics}
\usepackage{IEEEeqnarray,progSTF}
\usepackage{listings}
\usepackage{float}
\usepackage{epsfig}
\usepackage{pstricks,graphics}
\usepackage{enumerate}
\usepackage{caption}
\usepackage{amsmath,amssymb}
\usepackage{balance}
\usepackage{amsmath,amssymb}
\usepackage{alltt,url}
\usepackage{xspace}
\usepackage{epsfig}
\usepackage{pstricks,graphics}

\usepackage{balance}

\usepackage{amsmath,amsfonts,amsbsy} %, stmaryrd}
\usepackage{fancyvrb,keyval,ifthen}
\usepackage{graphicx}

\newtheorem{lemma}{Lemma}
\newtheorem{theorem}{Theorem}
\newtheorem{definition}{Definition}

\newcommand\proof{\noindent\textit{Proof.}~}

\newcommand\myendproof{\xspace$\Box$\\\indent}

\newcommand\hk{Hong \& Kung\xspace}

 \def\deltaI{\textit{dI}}
 \def\deltaO{\textit{dO}}
\newcommand{\vctr}[1]{\vec{#1}}
\newcommand{\mat}[1]{\mathbf{#1}}

\def\C{\textit{C}\xspace} 
\def\P{{\cal P}\xspace}
\def\Q{\textit{Q}\xspace}

\def\S{\textit{S}\xspace} 
\def\U{\textit{U}\xspace} 
\def\LB{\textit{L}\xspace}

\def\IO{\textit{IO}\xspace} 
\newcommand\Hmin[1]{\textit{H}\xspace}

\def\V{V\xspace}

\newcommand\card[1]{|#1|}
\newcommand\Min[1]{\textsf{Min}(#1)}
\newcommand\In[1]{\textsf{In}(#1)}
\newcommand\Out[1]{\textsf{Out}(#1)}

\newcommand\Doms[1]{\textsf{Dom}(#1)}

\newcommand*{\rom}[1]{\uppercase\expandafter{\romannumeral #1\relax}}

\long\def\comment#1{}
\begin{document}

\setlength{\pdfpageheight}{\paperheight}
\setlength{\pdfpagewidth}{\paperwidth}

\exclusivelicense
\conferenceinfo{POPL~'15}{January 15--17, 2015, Mumbai, India}
\copyrightyear{2015}
\copyrightdata{978-1-4503-3300-9/15/01}
\doi{2676726.2677010}

\title{On Characterizing the Data Access Complexity of Programs}
%\authorinfo{\small -- Submitted for anonymous review to POPL'15 --}{}{}
\authorinfo{Venmugil~Elango}
{The Ohio State University}
{elango.4@osu.edu}
\authorinfo
{Fabrice~Rastello}
{Inria}
{Fabrice.Rastello@inria.fr}
\authorinfo
{Louis-No\"el~Pouchet}
{The Ohio State University}
{pouchet@cse.ohio-state.edu}
\authorinfo
{J.~Ramanujam}
{Louisiana State University}
{ram@cct.lsu.edu}
\authorinfo
{P.~Sadayappan}
{The Ohio State University}
{saday@cse.ohio-state.edu}
%\authorinfo{}{}{}
\maketitle

\begin{abstract}
\input{abstract}
\end{abstract}

%% \category{CR-number}{subcategory}{third-level}

%% \terms
%% term1, term2

%% \keywords
%% keyword1, keyword2

\input{introduction}
\input{background}

\input{decomposition}
\input{rbw}

\input{static-analysis-lb}
%\input{case-studies-lb}
\input{related}
\input{disc}
\input{conclusion}

\section*{Acknowledgment}
We thank the anonymous referees for the feedback and many suggestions
that helped us significantly in
improving the presentation of the work. We thank Gianfranco Bilardi,
Jim Demmel, and Nick Knight for discussions on many aspects of lower bounds
modeling and their suggestions for improving the paper.
This work was supported in part by the U.S. National Science
Foundation through awards 0811457, 0926127, 0926687 and 1059417,
by the U.S. Department of Energy through award
DE-SC0012489, and by Louisiana State University.

%This research was supported in part by award xxx from the U.S. National
%Science Foundation and awards yyy and zzz from
%the U.S. Department of Energy.

%\newpage
%\appendix
%\input{appendix}

\balance
\bibliographystyle{abbrvnat}
\bibliography{iobib}

\end{document}

%% file: abstract.tex
Technology trends will cause data movement to account for the majority
of energy expenditure and execution time on emerging computers.
Therefore, computational complexity will no longer be a sufficient
metric for comparing algorithms, and a fundamental characterization of
data access complexity will be increasingly important.  The problem of
developing lower bounds for data access complexity has been modeled
using the formalism of Hong \& Kung's red/blue pebble game for
computational directed acyclic graphs (CDAGs).  However, previously
developed approaches to lower bounds analysis for the red/blue pebble
game are very limited in effectiveness when applied to CDAGs of real
programs, with computations comprised of multiple sub-computations
with differing DAG structure. We address this problem by developing an
approach for effectively composing lower bounds based on graph
decomposition. We also develop a static analysis algorithm to
derive the asymptotic
data-access lower bounds of programs, as a function of the
problem size and cache size.  %As an illustration of
%the use of the developments, we present several case studies, along
%with the first known lower bounds analysis of the data movement cost
%of a member of the family of recently developed ``communication
%avoiding'' Krylov subspace methods for iterative solution of sparse
%linear systems.

%% file: introduction.tex
%\vspace*{-6ex}
\section{Introduction}
%\vspace*{-2ex}
Advances in technology over the last few decades have yielded
significantly different rates of improvement in the computational
performance of processors relative to the speed of memory access.
Because of the significant mismatch between computational latency and
throughput when compared to main memory latency and bandwidth, the use
of hierarchical memory systems and the exploitation of significant
data reuse in the faster (i.e., higher) levels of the memory hierarchy
is critical for high performance. 
With future systems, the cost of data movement through the memory hierarchy
is expected to become even more dominant relative to the cost of
performing arithmetic operations \cite{bergman2008exascale,fuller2011,shalf2011exascale}, both in terms of
time and energy.
It is therefore of critical importance to limit the
  volume of data movement to/from memory by enhancing data reuse in
  registers and higher levels of the cache.
Thus the characterization of the inherent data access complexity of
computations is extremely important. 

\newsavebox{\firstlisting}
\begin{lrbox}{\firstlisting}
\lstset{language=C,basicstyle=\ttfamily}
{\small
\begin{lstlisting}
for(i=1;i<N-1;i++)
  for(j=1;j<N-1;j++)
    A[i,j]=A[i-1,j]+A[i,j-1];
\end{lstlisting}
}
\end{lrbox}

\newsavebox{\secondlisting}
\begin{lrbox}{\secondlisting}
\lstset{language=C,basicstyle=\ttfamily}
{\small
\begin{lstlisting}
for(it=1;it<N-1;it+=T)
  for(jt=1;jt<N-1;jt+=T)
    for(i=it;i<min(it+T,N-1);i++)
      for(j=jt;j<min(jt+T,N-1);j++)
        A[i,j]=A[i-1,j]+A[i,j-1];
\end{lstlisting}
}
\end{lrbox}

\begin{figure}[h!tb]
\centering
%\addtolength{\subfigcapskip}{1em}
\subfigure[Untiled code] {\usebox{\firstlisting}} 
\subfigure[Equivalent tiled code] {\usebox{\secondlisting}} 
%\subfigure[CDAG] {\raisebox{-1cm}{\includegraphics[width=3.0cm,viewport = 0 0 200 200]{./figures/seidel_cdag.eps}}}
\subfigure[CDAG] {\includegraphics[width=3.0cm,viewport = 0 0 200 200]{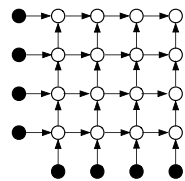}}
\caption{\label{fig:Seidel} Single-sweep two-point Gauss-Seidel code}
\end{figure}
%\lstset{language=C,basicstyle=\small\ttfamily}
%\begin{figure}[h!tb]
%\begin{center}
%\begin{minipage}{.99\columnwidth}
%\begin{minipage}{.59\columnwidth}
%\begin{lstlisting}
%for(i=1;i<N-1;i++)
%  for(j=1;j<N-1;j++)
%    A[i,j]=A[i-1,j]+A[i,j-1];
%\end{lstlisting}
%\captionof{figure}{(a) Untiled code}
%\begin{lstlisting}
%for(it=1;it<N-1;it+=T)
%  for(jt=1;jt<N-1;jt+=T)
%    for(i=it;i<min(it+T,N-1);i++)
%      for(j=jt;j<min(jt+T,N-1);j++)
%        A[i,j]=A[i-1,j]+A[i,j-1];
%\end{lstlisting}
%\captionof{figure}{(b) Equivalent tiled code}
%\end{minipage}
%~~~
%\begin{minipage}{.38\columnwidth}
%\includegraphics[width=\columnwidth]{./figures/seidel_cdag.eps}
%\end{minipage}
%\end{minipage}
%\end{center}
%\caption{\label{fig:Seidel} Single-sweep two-point Gauss-Seidel code}
%\end{figure}

Let us consider the code shown in Fig.~\ref{fig:Seidel}(a). Its computational
complexity can be simply stated as $(N-2)^2$ arithmetic
operations.  Fig.~\ref{fig:Seidel}(b) shows a functionally equivalent form of the
same computation, after a tiling transformation. The tiled form too has
exactly the same computational complexity of $(N-2)^2$ arithmetic
operations. 
Next, let us consider the data access cost for execution of these two
code forms on a processor with a single level of cache.
If the problem size $N$ is larger than cache capacity, the number of
cache misses would be higher for the untiled version
(Fig.~\ref{fig:Seidel}(a)) than the tiled version
(Fig.~\ref{fig:Seidel}(b)). But if the cache size were sufficiently
large, the tiled version would not offer any
benefits in reducing cache misses.

Thus, unlike the computational complexity of an algorithm, which stays
unchanged for different valid orders of execution of its operations
and also independent of machine parameters like cache size, the data
access cost depends both on the cache capacity and the order of
execution of the operations of the algorithm.

A fundamental question therefore is: {\em Given a computation 
  and the amount of storage at different levels of the cache/memory hierarchy,
  what is the
  minimum possible number of data transfers at the different levels,
  among all valid schedules that perform the operations?} 

In order to model the range of valid scheduling orders for the operations
of an algorithm, it is common to use the abstraction of the
computational directed acyclic
graph (CDAG), with a vertex for each instance of each computational
operation, and edges from producer instances to consumer instances.
Fig.~\ref{fig:Seidel}(c) shows the CDAG for the 
codes in Fig.~\ref{fig:Seidel}(a) and Fig.~\ref{fig:Seidel}(b), for $N$=6;
although the relative order of operations is different between the tiled and untiled
versions, the set of computation instances and the producer-consumer relationships
for the flow of data are exactly the same
(special ``input'' vertices in the CDAG represent values of elements of $A$ that
are read before they are written in the nested loop).

While in general it is intractable to precisely answer the above fundamental
question on the absolute minimum number of data transfers between main
memory and caches/registers among all valid execution schedules of a CDAG,
it is feasible to develop
{\em lower bounds} on the optimal number of data transfers.

An approach to developing a lower bound on the minimal data movement 
for a computation 
in a two-level memory hierarchy
was addressed in the seminal work of \hk 
%(where it was called I/O complexity; we will use both terms
%interchangeably)
by
using the model of the red/blue pebble game on a computational
directed acyclic graph (CDAG) \cite{hong.81.stoc}. While the approach
has been used to develop I/O lower bounds for a small number of 
homogeneous computational kernels, as elaborated later, it poses challenges
for effective analysis of full applications that are comprised of a number
of parts with differing CDAG structure.

In this paper, we address the problem of analysis of
affine loop programs to develop lower
bounds on their data movement complexity. The work presented in
this paper makes the following contributions:
\begin{compactitem}
\item {\bf Enabling composition in analysis of data access lower bounds}:
It adapts
the Hong \& Kung pebble game model on CDAGs
and the associated model of {\em S-partitioning} under a restriction
that disallows recomputation,
thereby enabling effective composition of I/O lower bounds for composite CDAGs
from lower bounds for component CDAGs.
%While such a restriction has also been previously considered by others, we
%are the first to formalize effective composition rules and
%decomposition strategies under such a model, critical to the analysis
%of general programs.
\item {\bf Static analysis of programs for lower bounds characterization}:
It develops an approach for asymptotic parametric analysis of data-access
lower bounds for arbitrary affine loop programs, as a function of cache size 
and problem size.
This is done by analyzing linearly independent families
of non-intersecting dependence chains.
%We use several case studies to demonstrate the use
%of the developed approach, including the first (to the best of
%our knowledge) lower
%bounds analysis for a member of the recently proposed 
%``communication avoiding'' iterative sparse linear system solvers
%using Krylov subspace methods\cite{hoemmen2010communication}.
\end{compactitem}

%% file: background.tex
\section{Background}
\label{sec:background}
\subsection{Computational Model}
We are interested in modeling the inherent data access complexity of a
computation, defined as the minimum number of data elements to be moved between
local memory (with limited capacity but fast access by the processor)
and main memory (much slower access but unlimited capacity)
among all valid execution orders for the operations making up
the computation. While the key developments in this paper
can be naturally extended to address multi-level
memory hierarchies and parallel execution, using an approach like
the MMHG (Multiprocessor
Memory Hierarchy Game) model of Savage 
\& Zubair~\cite{savage2008unified},
we restrict the treatment in this paper
to the case of only two levels of memory hierarchy and sequential
execution.

\begin{figure}[h!tb]
%\begin{figure}[t]
\begin{center}
\begin{minipage}{4cm}
\begin{verbatim}
for (i = 1; i < 4; ++i)
  S += A[i-1] + A[i];
\end{verbatim}
\hrule
\end{minipage}

\includegraphics[width=4cm]{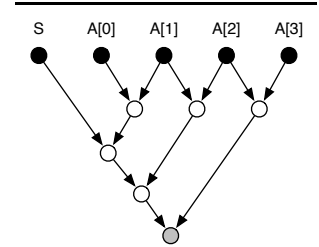}
\end{center}
\caption{\label{fig:ex-cdag} Example of a CDAG. Input vertices are represented in black, output vertices in grey.}
\end{figure}

The model of computation we use is a computational directed acyclic
graph (CDAG), where computational operations are represented as graph
vertices and the flow of values between operations is captured by
graph edges.  
Fig.~\ref{fig:ex-cdag} shows an example of a CDAG
corresponding to a simple loop program. 
Two important characteristics
of this abstract form of representing a computation are that (1) there
is no specification of a particular order of execution of the
operations: although the program executes the operations in a
specific sequential order, the CDAG abstracts the schedule of
operations by only specifying partial ordering constraints as edges in
the graph; (2) there is no association of memory locations with the
source operands or result of any operation.
(labels in
Fig.~\ref{fig:ex-cdag} are only shown for aiding explanation; they are not part of
the formal description of a CDAG).

\noindent We use the notation of
Bilardi \& Peserico~\cite{bilardi2001characterization} to formally describe
the CDAG model used by \hk:
%\vspace*{-1ex}
%{\small \begin{definition}[CDAG-HK]
{\begin{definition}[CDAG-HK]
A computational directed acyclic graph (CDAG) is a 4-tuple
$C = (I,V,E,O)$ of finite sets such that:
(1) $I \subset V$ is the input set and all its vertices have no incoming edges;
(2) $E \subseteq V \times V$ is the set of edges;
(3) $G = (V, E)$ is a directed acyclic graph;
% with no isolated vertices;
%FAB: removed from here. Added in Def 2 (RB pebble game)
(4) $V\setminus I$ is called the operation set and all its vertices have one
or more incoming edges;
(5) $O \subseteq V$ is called the output set.
\end{definition}}

\begin{figure}
\centering
%\begin{minipage}{4cm}
%\begin{verbatim}
%for (i = 1; i < 4; ++i)
%%  S += A[i-1] + A[i];
%\end{verbatim}
%\hrule
%\end{minipage}
\includegraphics[width=4cm]{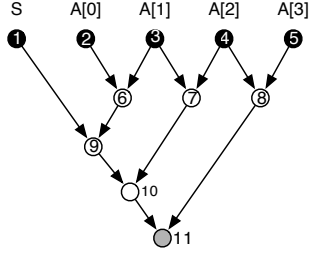}
\caption{\label{fig:ex-pebblegame} Example of schedule for a {\em
complete calculation} on CDAG in Fig.~\ref{fig:ex-cdag}. The vertex
numbers represent the order of execution.}
\end{figure}

%FAB: H&K requires I and O to be disjoint because they do not have white pebble to impose all nodes to be computed at least once. We do not but the pb is unchanged if I is removed from O. So we consider O and I to be disjoint.
%
%%FAB: H&K need that any node is an ancestor of some output node; In the construction of the 2S-partitioning (from a valid game) the dom they choose is made only of some pred nodes.
%%FAB: Check why we would need this. Not good for the decomposition. Might be necessary for the RB pebble game but not for the red-blue-white (as at the end every game every body must have a white).
%% In this paper, we use the above definition of the CDAG from Bilardi and
%% Peserico \cite{bilardi2001characterization} along with an additional
%% constraint: every vertex is reachable from at least one input vertex
%% and at least one output vertex is reachable from every vertex. Also
%For convenience, we denote $\V=V\cup I$.
%
%% We next summarize the pebbling model that has been used
%% to characterize the minimal data
%% access complexity (traditionally called I/O complexity) of a
%% CDAG on a processor with limited ``fast'' memory.
%
\subsection{The Red-Blue Pebble Game}
\label{sec:rbpg}
\noindent \hk used this computational model in their seminal work
\cite{hong.81.stoc}.  The inherent I/O complexity of a CDAG is the
minimal number of I/O operations needed while optimally playing the
\emph{Red-Blue pebble game}. This game uses two kinds of
pebbles: a fixed number of red pebbles that represent the small fast
local memory (could represent cache, registers, etc.), and an
arbitrarily large number of blue pebbles that represent the large slow
main memory. 
%%% this text is repeated in the defn below //Ram
%%% Starting with blue pebbles on all inputs nodes in the CDAG, 
%%% a complete calculation 
%%% involves the generation of a sequence of steps to finally produce
%%% blue pebbles on all outputs. 
%
%\vspace*{-1ex}
%{\small \begin{definition}[Red-Blue pebble game~\cite{hong.81.stoc}]
{\begin{definition}[Red-Blue pebble game~\cite{hong.81.stoc}]
\label{def:rbpg}
Let $C=(I,V,E,O)$ be a CDAG such that any vertex with 
no incoming (resp. outgoing) edge is an element of $I$ (resp. $O$).
Given \S red pebbles and an arbitrary number
of blue pebbles, with an initial blue pebble on each \textit{input} vertex,
a complete calculation is any sequence of steps using the following
rules that results in
a final configuration with blue pebbles on all \textit{output} vertices:
%\vspace*{-1ex}
\begin{description}
%\begin{tightlist}
\item[\textbf{R1 (Input)}] A red pebble may be placed on any vertex that has a blue pebble (load from slow to fast memory),
\item[\textbf{R2 (Output)}] A blue pebble may be placed on any vertex
that has a red pebble (store from fast to slow memory),
\item[\textbf{R3 (Compute)}] If all immediate predecessors of a vertex
	$v \in V\setminus I$
%~\footnote{The original red-blue pebble game described in~\cite{hong.81.stoc} allows incorrectly the computation of an input node: the proof of Theorem~3.1 assumes the set of predecessors of a computational vertex $v$ to be a dominator set. For any input node it would lead to an empty dominator set instead of $\{v\}$ itself.} 
have red pebbles, a red pebble may be placed on (or moved
to)~\footnote{The original red-blue pebble game in \cite{hong.81.stoc}
does not allow moving/sliding a red pebble from a predecessor vertex to a successor; we chose to allow it since
it reflects real instruction set architectures. Others \cite{savage.cc.95} have
also considered a similar modification. But all our proofs hold for both
the variants.}
$v$ (execution or ``firing'' of operation),
\item[\textbf{R4 (Delete)}] A red pebble may be removed from any vertex (reuse storage).
\end{description}
%\end{tightlist}
%\vspace*{-2ex}
\end{definition}
}%
\noindent The number of I/O operations for any complete calculation is the
total number of moves using rules R1 or R2, i.e., the total number
of data movements between the fast and slow memories. The inherent I/O
complexity of a CDAG is the smallest number of such I/O operations
that can be achieved, among all complete calculations
for that CDAG. An \emph{optimal} calculation is a complete calculation
achieving the minimum  number of I/O operations.

Fig.~\ref{fig:ex-pebblegame} shows an example schedule
for the CDAG
in Fig.~\ref{fig:ex-cdag}.
%are used simply to the show the correspondence between the loop code
%and its CDAG, but are not part of
%the formal description of the CDAG).
Given $\S$ red pebbles and unlimited blue pebbles, goal of the game is to begin with blue pebbles on all input
vertices, and finish with blue pebbles on all output vertices by following the rules
in Definition \ref{def:rbpg} without using more than $\S$ red pebbles. Considering the case with three red pebbles ($\S=3$),
one possible complete calculation for the CDAG in Fig. \ref{fig:ex-pebblegame} is:
\{$R1_2$, $R1_3$, $R3_6$, $R4_2$, $R1_1$, $R3_9$, $R4_1$, $R4_6$,
$R1_4$, $R3_7$, $R4_3$, $R3_{10}$, $R4_9$, $R4_7$, $R1_5$, $R3_8$,
$R4_4$, $R4_5$, $R3_{11}$, $R2_{11}$\}.
%~\footnote{The vertex numbers are shown
%in the subscript for easy reference; but are not part of the complete
%calculation.}. 
The I/O cost of this
complete calculation is 6 (which corresponds to the number of moves
using rules $R1$ and $R2$). A different complete calculation for the
same CDAG with I/O
cost of 12 is given by: \{$R1_2$, $R1_3$, $R3_6$, $R4_2$, $R1_4$, $R2_6$, $R3_7$,
$R4_3$, $R1_5$, $R2_7$, $R3_8$, $R2_8$, $R1_1$, $R1_6$, $R3_9$,
$R4_1$, $R4_6$, $R1_7$, $R3_{10}$, $R4_7$, $R4_9$, $R1_8$, $R3_{11}$,
$R2_{11}$\}. The I/O complexity of the CDAG is the minimum I/O cost
of all such complete calculations.

\subsection{Lower Bounds on I/O Complexity via \S-Partitioning}
While the red-blue pebble game provides an operational definition for the
I/O complexity problem, it is generally not feasible to
determine an optimal calculation on a CDAG.  \hk developed a
novel approach for deriving I/O lower bounds for CDAGs by
relating the red-blue pebble game to a graph partitioning problem
defined as follows.
%
%\vspace*{-1ex}
%{\small \begin{definition}[\hk \S-partitioning of a CDAG~\cite{hong.81.stoc}]
{\begin{definition}[\hk \S-partitioning of a CDAG~\cite{hong.81.stoc}]
\label{def:spart}
Let $\C=(I,V,E,O)$ be a CDAG. An \S-partitioning of $\C$ is a
collection of $h$ subsets of $\V$ such that:
%\begin{compactenum}
\begin{description}
%\item[\textbf{(P1)}] $\bigcap_{i = 1}^h \V_i = \emptyset$, and $\bigcup_{i = 1}^h \V_i = \V$
\item[\textbf{P1}] $\forall i\neq j,\ \V_i\cap \V_j = \emptyset$, and $\bigcup_{i = 1}^h \V_i = \V$
%\item[\textbf{(P2)}] there is no circuit between subsets
\item[\textbf{P2}] there is no cyclic dependence between subsets
\item[\textbf{P3}] $\forall i,~~\exists D\in \Doms{{\V}_i}~~\textrm{such that}~~\card{D} \le \S$
\item[\textbf{P4}] $\forall i,~~\card{\Min{{\V}_i}} \le \S$
\end{description}
%\end{compactenum}
\noindent where a dominator set of $\V_i$, $D\in\Doms{{\V}_i}$ is a set of vertices such that any path from $I$ to a vertex in $\V_i$ contains some vertex in $D$; the minimum set of $\V_i$, $\Min{{\V}_i}$ is the set of vertices in
${\V}_i$ that have all its successors outside of ${\V}_i$; 
and for a set $\textit{A}$, $\card{\textit{A}}$ is the
cardinality of the set $\textit{A}$.
%We say that there is a circuit between two sets $V_i$ and $V_j$,
   %if there is an edge from some vertex in $V_i$ to a vertex in $V_j$
   %and vice-versa.
\end{definition}
}
\hk showed a construction for a 2\S-partition of a CDAG,
corresponding to any complete calculation on that CDAG using
\S red pebbles, with a tight relationship between the number of vertex sets
$h$ in the 2\S-partition and the number of I/O moves $q$ in the
complete calculation, as shown in Theorem~\ref{thm.hk}. The tight
association between any complete calculation and a
corresponding 2\S-partition provides the key Lemma~\ref{lemma:hk} that
serves as the basis for \hk's approach for deriving lower bounds on the
I/O complexity of CDAGs typically by reasoning on the maximal number
of vertices that could belong to any vertex-set in a valid
2\S-partition.

\begin{theorem}[Pebble game, I/O and 2\S-partition~\cite{hong.81.stoc}]\label{thm.hk} %[Relation between I/O of pebble game and subsets in 2\S-partition~\cite{hong.81.stoc}]
Any complete calculation of the red-blue pebble game on a CDAG
using at most \S red pebbles is associated with a 2\S-partition of the CDAG
such that 
%$ \S\times h \ge q \ge \S\times(h-1), $
$ \S \; h \ge q \ge \S \; (h-1), $
where $q$ is the number of I/O moves in the complete calculation 
and $h$ is the number of subsets in the 2\S-partition.
\end{theorem}
%
%
%%%%%%%%%%%%%%%%%%%%%%%%%
%%%%%%%%%%%%%%%%%%%%%%%%%
\input{./fig_decomposition}
%%%%%%%%%%%%%%%%%%%%%%%%%
%%%%%%%%%%%%%%%%%%%%%%%%%
%
%
%
%\vspace*{-1ex}
%\vspace{-.55cm}
\begin{lemma}[Lower bound on I/O~\cite{hong.81.stoc}]
\label{lemma:hk}
Let $\Hmin{2\S}$ be the minimal number of vertex sets for any valid
$2\S$-partition of a given CDAG (such that any vertex with no incoming -- resp. outgoing -- edge is an element of $I$ -- resp. $O$). Then the minimal number \Q
of I/O operations for any complete calculation on the CDAG is bounded by: 
$\Q \ge \S \times(\Hmin{2\S}-1)$
\end{lemma}
%\vspace{-.25cm}
%
This key lemma has been useful in proving I/O lower bounds for several
CDAGs~\cite{hong.81.stoc} by reasoning about the maximal number of
vertices that could belong to any vertex-set in a valid 2\S-partition.

%% file: fig_decomposition.tex
\begin{figure*}[h!tb]
\centering
\begin{minipage}[b]{.95\textwidth}
\begin{minipage}[b]{0.31\textwidth}
\begin{verbatim}
for(i = 0; i < 4; i++) 
  c[i] = a[i] + b[i]; // S1
for(i = 0; i < 4; i++) 
  d[i] = c[i] * c[i]; // S2
for(i = 0; i < 4; i++) 
  e[i] = c[i] + d[i]; // S3
for(i = 0; i < 4; i++) 
  f[i] = d[i] * e[i]; // S4
\end{verbatim}

\centering (a) Original code
\end{minipage}
~
\begin{minipage}[b]{0.31\textwidth}
\hspace{-.5cm}\includegraphics[width=6cm]{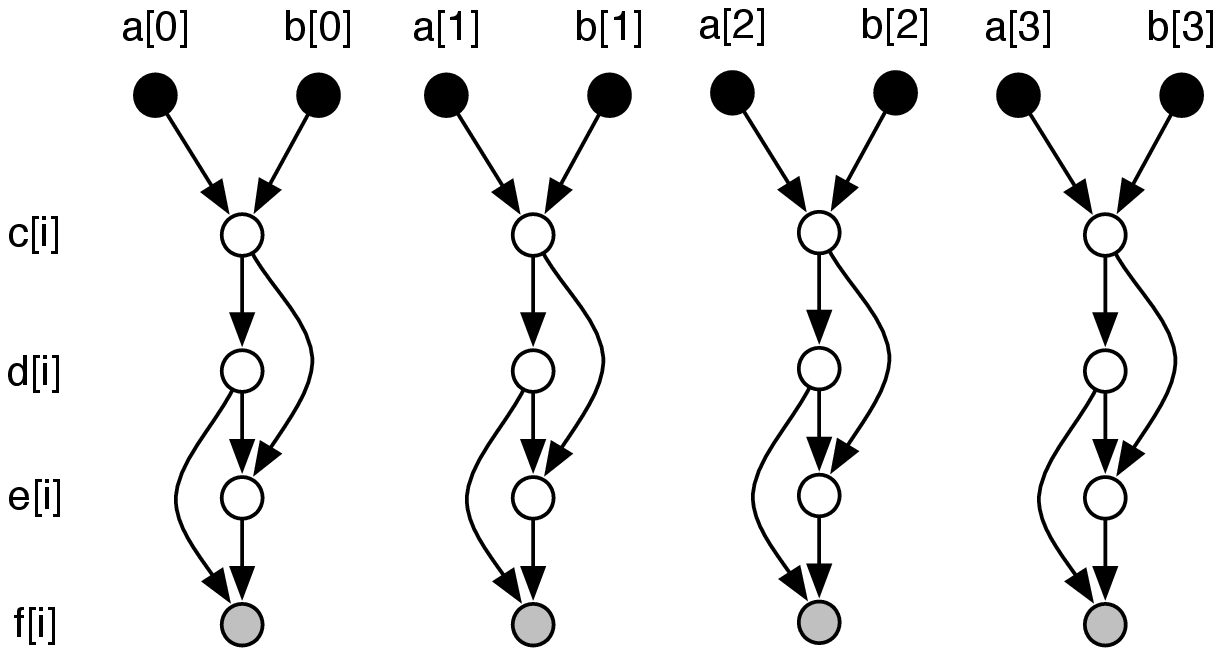}

\centering (b) Full CDAG
\end{minipage}
~
\begin{minipage}[b]{0.31\textwidth}
\hspace{.5cm}\includegraphics[width=6cm]{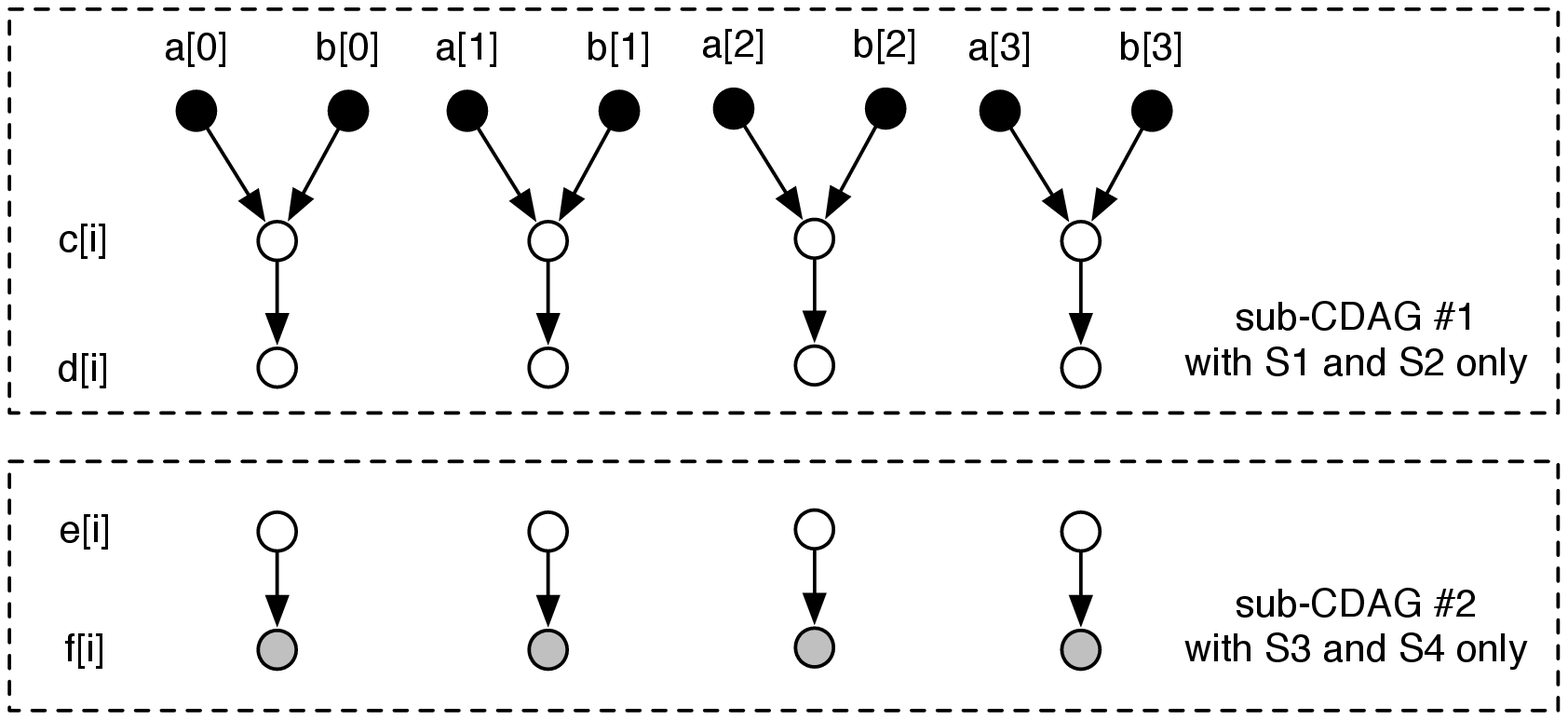}

\centering (c) CDAG partitioning
\end{minipage}
\end{minipage}
\caption{\label{fig:cdag-4stmts} Example illustrating limitation of \hk model regarding composition of lower bounds from sub-components of CDAG }
\end{figure*}

%% file: decomposition.tex
%\vspace*{-2ex}
\section{Challenges in Composing I/O Lower Bounds from Partitioned CDAGs}
\label{sec:decomposition}
%\vspace*{-2ex}
%\input{./fig_decomposition}

Application codes are typically constructed from a number of
sub-computations using the fundamental composition mechanisms of
sequencing, iteration and recursion. As explained in Sec. 1,
in contrast to analysis of
computational complexity of such composite application codes, I/O
complexity analysis poses challenges.  With computational complexity,
the operation counts of sub-computations can simply be added.
However, using the
red/blue pebble game model of \hk, as elaborated below, it is 
problematic to analyze the I/O complexity of sub-computations and simply
combine them by addition. In the next section, we develop an approach
to overcome the problem.

%\subsection{Overview of the Problem and Solution Approach}
%\label{sec:decomp-prob}

\subsection{\bf The Decomposition Problem} 
The \hk red/blue pebble game model places blue pebbles on all CDAG vertices
without predecessors, since such vertices are considered to hold
inputs to the computation, and therefore assumed to start off in slow memory.
Similarly, all vertices without successors are considered to be outputs
of the computation, and must have blue pebbles at the end of a complete calculation.
If the vertices of a CDAG corresponding to a composite application are
disjointly partitioned
into sub-DAGs, the analysis of each sub-DAG
will require the initial placement
of blue pebbles on all vertices without predecessors in the sub-DAG, and final
placement of blue pebbles on all vertices without successors in the sub-DAG.
So an optimal calculation for each sub-DAG will require at least one load
(R1) operation for each input and a store (R2) operation for each output.
But in a complete calculation on the full composite CDAG, clearly
it may be possible to pass values in a red pebble between vertices
in different sub-DAGs, so that the I/O complexity could be less than the sum
of the I/O costs for optimal calculations on each sub-DAG.
This is illustrated by the following example.

Fig.~\ref{fig:cdag-4stmts}(b) shows the CDAG for the computation in
Fig.~\ref{fig:cdag-4stmts}(a).  Fig.~\ref{fig:cdag-4stmts}(c) shows
the CDAG partitioned into two sub-DAGs, where the first sub-DAG
contains vertices of $S1$ and $S2$ (and the input vertices corresponding to
\texttt{a[i]} and \texttt{b[i]}), and the second sub-DAG contains
vertices of $S3$ and $S4$.
Considering the full CDAG, with just two red pebbles, it can be
computed at an I/O cost of $12$, incurring I/O just for the initial
loads of inputs \texttt{a[i]} and \texttt{b[i]}, and the final stores for outputs
\texttt{f[i]}. In contrast, with the partitioned sub-DAGs, the first sub-DAG
will incur additional output stores for the successor-free vertices
$S2[i]$, and the second sub-DAG will incur input loads for
predecessor-free vertices $S3[i]$. Thus the sum of optimal red/blue
pebble game I/O costs for the two sub-DAGs amounts to $20$ moves, i.e.,
it exceeds the optimal I/O cost for the full CDAG.

The above example illustrates a fundamental problem with the \hk
red/blue pebble game model: a simple combining of I/O lower
bounds for sub-DAGs of a CDAG cannot be used to generate an
I/O lower bound for the composite CDAG. 
But the ability to perform complexity analysis by combining analyses
of component sub-computations is important for the analysis of
real applications. Such decomposition of
data-access complexity analysis can be enabled by making
a change to the \hk
pebble game model, as discussed next.

\subsection{Flexible Input/Output Vertex Labeling to Enable Composition of Lower Bounds} 
With the \hk
model, all vertices without predecessors must be input vertices, and
all vertices without successors must be output vertices. By relaxing this
constraint, we show that composition of lower bounds from sub-CDAGs is
valid.
With such a modification,
vertices without predecessors will not be required to
be input vertices, and such predecessor-free non-input vertices
do not have an initial blue pebble
placed on them. However, such vertices are allowed to fire using
rule R3 at any time, since they do not have any predecessor nodes
without red pebbles. Vertices without successors 
are similarly not required to be output vertices, and those not designated as
outputs do not need a blue pebble on them at the
end of the game. However, all compute vertices (i.e., vertices in
$V\setminus I)$ in CDAG are required
to have fired for any complete calculation.\\
%
%\noindent{\bf Composing I/O Lower Bounds by CDAG Decomposition}:
%\label{subsec:decomposition}
Using the modified model of the red/blue pebble game with flexible input/output
vertex labeling, it is feasible to compose I/O lower bounds by adding
lower bounds for disjointly partitioned sub-CDAGs of a CDAG.
The following theorem formalizes it.
%The complete proof can be found
%in appendix~\ref{ap:decomposition}.

\begin{theorem}[Decomposition]
\label{thm:decomposition}
%Let $C=(I,V,E,O)$ be a CDAG. Let $V_1, V_2, \dots, V_p$ be an arbitrary
%(not necessarily acyclic) disjoint partitioning of $V$ ($i\neq j
%\Rightarrow V_i\cap V_j=\emptyset$ and $\bigcup_{1\leq i\leq p} V_i=V$)
%and $C_1, C_2, \dots, C_p$ be the induced partitioning of $C$
%($I_i=I\cap V_i$, $E_i=E\cap V_i\times V_i$, $O_i=O\cap V_i$).  Then
%$\sum_{1\leq i\leq p} \IO(C_i)\leq \IO(C)$. In particular, if $Q_i$ is a
%lower bound on the \IO of $C_i$, then $\sum_{1\leq i\leq p} Q_i$ is a
%lower bound on the I/O of $C$.
Let $C=(I,V,E,O)$ be a CDAG. Let $\{V_1, V_2, \dots, V_p\}$ be an arbitrary
(not necessarily acyclic) disjoint partitioning of $V$ 
%($i\neq j
%\Rightarrow V_i\cap V_j=\emptyset$ and $\bigcup_{1\leq i\leq p} V_i=V$)
($\bigcap_{i=1}^p V_i=\emptyset$ and $\bigcup_{i=1}^p V_i=V$)
and $C_1, C_2, \dots, C_p$ be the induced partitioning of $C$
($I_i=I\cap V_i$, $E_i=E\cap V_i\times V_i$, $O_i=O\cap V_i$).
If $\Q$ is the I/O complexity for $C$ and $\Q_i$ is the I/O complexity
for $C_i$, then $\sum_{i=1}^p \Q_i \le \Q$. In particular,
if ${\LB}_i$ is the I/O lower
bound for $C_i$, then
$\sum_{i=1}^p {\LB}_i$ is an I/O
lower bound for $C$.
\end{theorem}

\proof
Consider an optimal calculation $\cal P$ for $C$, with cost $\Q$. We
define the cost of $\cal P$ restricted to $V_i$, denoted as $\Q_{|V_i}$, as the
number of $R1$ or $R2$ transitions in $\cal P$ that involve a vertex of $V_i$.
Clearly $\Q=\sum_{i=1}^p \Q_{|V_i}$. We will show that we can build from $\cal P$, a valid
complete calculation ${\cal P}_{|V_i}$ for $C_i$, of cost $\Q_{|V_i}$. This will prove that
$\Q_i\leq \Q_{|V_i}$, and thus  $\sum_{i=1}^p \Q_i\leq
\sum_{i=1}^p \Q_{|V_i} = \Q$. ${\cal P}_{|V_i}$ is built from
$\cal P$ as follows: (1) for any transition in $\cal P$ that involves a vertex
$v \in V_i$, apply this transition in  ${\cal P}_{|V_i}$; (2) delete all other
transitions in $\cal P$. Conditions for transitions $R1$, $R2$, and $R4$ are
trivially satisfied. Whenever a transition $R3$ on a vertex $v$ is performed in
$\cal P$, all the predecessors of $v$ must have a red pebble on them. Since all
transitions of $\cal P$ on the vertices of $V_i$ are maintained in ${\cal
P}_{|V_i}$, when $v$ is executed in ${\cal P}_{|V_i}$, all its
predecessor vertices must have red pebbles, enabling transition $R3$.
\myendproof

With this modified model of the red/blue pebble game that permits
predecessor-free vertices to be non-input vertices, complex CDAGs
can be decomposed and lower bounds for the composite CDAG can be
obtained by composition of the bounds from the sub-CDAGs. However,
sub-CDAGs that have no ``true'' input and output vertices in them will
have trivial I/O lower bounds of zero -- the entire set of
vertices in the sub-CDAG can fit in a single vertex set for a 
valid 2S-partition, for any value of S, since conditions {\bf P1-P4}
are trivially satisfied. 

In the next section, we present a solution to the problem.
The main idea is to impose restrictions on
the red/blue pebble game to disallow re-pebbling or multiple firings
of any vertex using rule R3. We show that by imposing such a restriction,
we can develop an input/output tagging strategy for sub-CDAGs that
enables stronger lower bounds to be generated by CDAG decomposition.

%% file: rbw.tex
%\vspace*{-4ex}
\section{S-Partitioning when Re-Pebbling is Prohibited}
\label{sec:rbwpg}
%\vspace*{-2ex}

With the pebble game model of \hk, the compute rule R3 could be applied
multiple times in a complete calculation. This is useful in modeling algorithms
that perform re-computation of multiply used values rather 
than incur the overhead of storing and loading it. 
%For example, some
%implementations of coupled cluster methods in quantum chemistry
%\cite{BartlettShavitt-book}
%perform ``on the fly'' regeneration of elements of integral tensors that
%are too large to fit in main memory and would be more expensive to
%store in disk and reload for each iteration in an iterative convergence
%loop. 
However, the majority of practically used algorithms do not
perform any redundant recomputation. Hence 
several efforts
  \cite{BDHS11,BDHS11a,bilardi2001characterization,bilardi12-bsp,michele13,ranjan11.fft,savage.cc.95,savage.book,savage.options,cook.pebbling,toledo.jpdc,ranjan12.rpyr,ranjan12.vertex}
have modeled I/O complexity under a more restrictive model that disallows
recomputation, primarily because it eases or enables analysis with some
lower bounding techniques. In this section, we consider the issue of
composing bounds via CDAG decomposition under a model that
disallows recomputation, i.e., prohibits re-pebbling. 
We develop a modified definition of S-partition that is adapted
to enable I/O lower bounds to be developed for the restricted 
red/blue pebble game. This provides two significant benefits:
\begin{compactenum}
\item It enables non-trivial I/O lower bound contributions to
be accumulated from sub-CDAGs of a CDAG, even when the sub-CDAGs
do not have any true inputs. This is achieved via input/output
tagging/untagging strategies we develop in this section.
\item It enables static analysis
of programs to develop parametric expressions for 
asymptotic lower bounds as a function of cache and
problem size parameters. This is described in the following 
sections.
\end{compactenum}
%We develop an
%approach to help answer the question: {\it Given an algorithm whose
%computational structure is captured as a CDAG, can it possibly benefit
%from an implementation that uses redundant recomputation to trade-off the
%extra operations for reduced data movement?}

%To address this question, we develop new lower bounding techniques for
%the red-blue pebble game when repebbling is disallowed. By developing
%lower bounds for a CDAG under both models -- the original \hk model
%that allows repebbling and a restricted model that prohibits repebbling --
%the question can be answered for many CDAGs.
%We show that it is possible to establish through tight lower bounds that
%some CDAGs have higher data movement requirements
%under a non-repebbling model, while for other CDAGs the lower bounds 
%under both models are the same. This implies that some computations have the
%potential to benefit from devising implementations that perform redundant
%computations to lower data movement costs, while other computations cannot
%benefit from such redundant recomputation.

A pebble game model that does not allow recomputation can be formalized
by changing rule R3 of the red/blue pebble game to R3-NR (NR denotes
No-Recomputation or No-Repebbling) and the definition of a
complete calculation as follows:
\begin{definition}[Recompute-restricted Red-Blue pebble game]
\label{def:rbwpg}
Let $C=(I,V,E,O)$ be a CDAG.
Given \S red pebbles and arbitrary number
of blue pebbles, with an initial blue pebble on each \textit{input} vertex,
a complete calculation is any sequence of steps using the following
rules that causes each vertex in $\V\setminus I$ to be fired once
using Rule R3-NR, and results in
a final configuration with blue pebbles on all \textit{output} vertices:
\begin{description}
\item[\textbf{R1 (Input)}] A red pebble may be placed on any vertex that has a blue pebble (load from slow to fast memory),
\item[\textbf{R2 (Output)}] A blue pebble may be placed on any vertex
that has a red pebble (store from fast to slow memory),
\item[\textbf{R3-NR (Compute)}] If all immediate predecessors of a vertex
	$v$ $\in$ $V\setminus I$ have red pebbles on them,
and a red pebble has not previously been placed
on $v$, a red pebble may be placed on $v$.
\item[\textbf{R4 (Delete)}] A red pebble may be removed from any vertex (reuse storage).
\end{description}
\end{definition}

We next present an adaptation of Hong and Kung's S-partition that will enable
us to develop larger lower bounds for the restricted pebble game model
that prohibits repebbling. 
%Clearly, since any valid game under this
%model is also a valid game for the standard red-blue game defined by \hk,
%the I/O lower bound for any CDAG for the standard model will be less than
%or equal to that under the restricted model. Hence, for some CDAGs, it
%may be possible to establish tight lower bounds for the restricted model
%that are higher than tight lower bounds for the standard model. Such CDAGs
%can potentially benefit from implementations that use recomputations. But
%if tight lower bounds under both models are the same, it implies that
%recomputation cannot benefit the computation with respect to a trade-off
%between extra operations and lower data movement.
%
%First, Definition~\ref{def:spart} is adapted to this new game so that
%Theorem~\ref{thm.hk} and thus Lemma~\ref{lemma:hk} can hold for the
%RBW pebble game.
%
\begin{definition}[$S^{NR}$-partitioning of CDAG]
%\begin{definition}[S-partitioning of CDAG]
\label{def:newspart}
Given a CDAG $\C$, an $S^{NR}$-partitioning of $\C$ is a
collection of $h$ subsets of $\V\setminus I$ such that:
\begin{description}
\item[\textbf{P1}] $\forall i\neq j,\ \V_i\cap \V_j = \emptyset$, and
	$\bigcup_{i = 1}^h V_i = \V\setminus I$
\item[\textbf{P2}] there is no cyclic dependence between subsets
\item[\textbf{P3}] $\forall i,~~\card{\In{{V}_i}} \le \S$
\item[\textbf{P4}] $\forall i,~~\card{\Out{{V}_i}} \le \S$
\end{description}
\noindent where the input set of $V_i$, $\In{{V}_i}$ is the set of
vertices of $\V\setminus V_i$ that have at least one successor in $V_i$; the
output set of $V_i$, $\Out{V_i}$ is the set of vertices of $V_i$ that
are also
part of the output set $O$ or that have at least one successor outside of ${\V}_i$.
%and $\card{\textit{Set}}$ is the cardinality of the set $\textit{Set}$.
\end{definition}
%
%We use the superscript $NR$, denoting ``No Repebbling" (or ``No Recompute"),
%for our adapted definition of
%S-partitioning that is well suited for developing I/O lower bounds for
%for the restricted game. The proof for the analog of Theorem~\ref{thm.hk} 
%for the restricted pebble game is provided in
%appendix~\ref{ap:thm.hk.rbw}.

\label{thm:thm.hk.rbw}
\begin{theorem}[Restricted pebble game, I/O and $2S^{NR}$-partition]
Any complete calculation of the red-blue pebble game, without repebbling, on a CDAG
using at most \S red pebbles is associated with a $2S^{NR}$-partition of the CDAG
such that $ \S\times h \ge q \ge \S\times(h-1), $
where $q$ is the number of I/O moves in the game and $h$ is the number of
subsets in the $2S^{NR}$-partition.
\end{theorem}
\proof
Consider a complete calculation $\P$ that corresponds to some scheduling
(i.e., execution)
of the vertices of the graph $G=(\V,E)$ that follows the rules R1--R4
of the restricted pebble game.
We view this calculation as a string that has recorded all
the transitions (applications of pebble game rules). Suppose that $\P$ contains exactly $q$ transitions of type $R1$ or $R2$.
Let $(\P_1,\P_2,\dots,\P_h)$ correspond to a partitioning of the transitions of $\P$ into
$h=\lceil q/\S\rceil$ consecutive sub-sequences such that each $\P_i
\in (\P_1,\dots,\P_{h-1})$ contains exactly $\S$ transitions of type $R1$ or $R2$.

The CDAG contains no node isolated from the output nodes, and any
vertex of $\V\setminus I$ is computed exactly once in $\P$.
Let $V_i$ be the set of vertices computed (transition R3-NR) in the
sub-calculation $\P_i$. Property $P1$ is trivially fulfilled.

As transition R3-NR on a vertex $v$ is possible only if its
predecessor vertices have red pebbles on them, those predecessors are
necessarily executed in some $\P_j$, $j\leq i$ and are thus part of a
$V_j$, $j\leq i$. This proves property $P2$.

To prove $P3$, for a given $V_i$ we consider two sets: $V_R$ is the set
of vertices that had a red pebble on them just before the execution of
$P_i$; $V_{BR}$ is the set of vertices on which a red pebble is placed
according to rule $R1$ (input) during $\P_i$. We have,
$\In{V_i}\subseteq V_R\cup V_{BR}$. Thus $\card{\In{V_i}}\leq
\card{V_R}+\card{V_{BR}}$. As there only
$\S$ red pebbles,  $\card{V_R}\leq \S$. Also by construction of $\P_i$,
$\card{V_{RB}}\leq \S$. This proves that $\card{\In{V_i}}\leq 2\S$
(property $P3$).

Property $P4$ is proved in a similar way: $V'_R$ is the set of
vertices that have a red pebble on them just after the execution of
$\P_i$; $V'_{RB}$ is the set of vertices of $V_i$ on which a blue
pebble is placed during $\P_i$ according to rule $R2$. We have that
$\Out{V_i}\subseteq V'_R\cup V'_{RB}$. Thus $\card{\Out{V_i}}\leq
\card{V'_R}+\card{V'_{RB}}$. As there are only
$\S$ red pebbles,  $\card{V'_R}\leq \S$. Also by construction of
$\P_i$, $\card{V'_{RB}}\leq \S$. This proves that $\card{\Out{V_i}}\leq
2\S$ (property $P4$).
\myendproof

\begin{lemma}[I/O lower bound for restricted pebble game]
\label{lemma:hk.rbw}
Let $H^{NR}$ be the minimal number of vertex sets for any valid
$2S^{NR}$-partition of a given CDAG.
Then the minimal number \Q
of I/O operations for any complete calculation on the CDAG, without
any repebbling, is bounded by:
$\Q \ge \S \times(H^{NR}-1)$
\end{lemma}

%graph. Such a refinement is needed to define proper decomposition rules,
%as shown below. We note that the modified S-partitioning from
%Def.~\ref{def:newspart} and the associated I/O complexity reasoning is
%also applicable to the flexible I/O RBW pebble game. For (sub-)graphs
%without input/output sets, the application of S-partitioning will
%however lead to a trivial partition with all vertices in a single set
%(e.g., $h = 1$). A careful tagging of vertices as virtual input/output
%nodes will be required for better I/O complexity estimates, as
%detailed below.

%Definition~\ref{def:rbwpg} allows us to partition a CDAG $\C$ into
%sub-CDAGs $C_1, C_2, \dots, C_p$, to compute lower bounds on the I/O
%complexity of each sub-CDAG $\IO(C_1), \IO(C_2), \dots, \IO(C_p)$
%independently and simply add them to bound the I/O complexity of
%$\C$. This is stated in the following decomposition theorem.

The above theorem and lemma establish the relationship between complete
calculations of the restricted pebble game and 2S-NR partitions.
The critical difference between the standard S-partition of \hk
and the S-NR partition is the validity condition pertaining to
incoming edges into a vertex set in the partition: for the former 
the size of
dominator sets is constrained to be no more than S, while for the latter 
the number of external vertices with edges into the vertex set
is constrained by S.
When a CDAG is decomposed into sub-CDAGs, very often some of the sub-CDAGs
get isolated from the CDAG's input and output vertices. 
%As illustrated by the R-MCL example in the previous section,
This will lead to trivial (i.e., zero) 
lower bounds for such component sub-CDAGs.
For the restricted pebble game, below
we develop an approach to obtain tighter lower bounds for component
sub-CDAGs that have become isolated from inputs and outputs of the full CDAG.
The key idea is to allow any vertex without predecessors (resp.~successors) 
to simulate an
input (resp.~output) vertex by specially tagging it so, 
and then adjusting the obtained
lower bound to account for a one-time access cost for loading (resp.~storing) such 
a tagged input (resp.~output).
%There are cases where separating input/output vertices leads to very weak lower
%bounds. This happens when input vertices have high fan out
%(Inequality~\ref{eq:insert} cannot hold) 
%such as for matrix-multiplication: if
%we consider the CDAG for matrix-multiplication and remove all input and output
%vertices, we get a set of independent chains that can each be computed with no
%more than 2 red pebbles. To overcome this problem, the following theorem
%allows us to compare the I/O of two CDAGs: a CDAG $C'=(I',V,E,O')$ and another
%$C=(I,V,E,O)$ built from $C'$ by just transforming some vertices without
%predecessors into input vertices, and some others into output nodes so that
%$I'\subset I$ and $O'\subset O$.
%In contrast to the prior development above,
%instead of adding/removing input/output vertices, here we do not change the
The vertices of a CDAG remain unchanged, but the labeling (tag) of some vertices
as inputs/outputs in the CDAG is changed.
%We state the theorem and
%provide a sketch of the proof below. Complete proof can be found in
%appendix~\ref{ap:io-compar}.
%So the DAG remains the same, but some
%input/output vertices are relabeled as standard computational vertices, or vice
%versa.
%
%\vspace{-.1cm}
\begin{theorem}[Input/Output (Un)Tagging -- Restricted pebble game]
\label{thm:io-compar}
Let $C$ and $C'$ be two CDAGs of the same DAG $G=(V,E)$: $C=(I,V,E,O)$,
$C'=(I\cup \deltaI,V,E,O\cup \deltaO)$, where, $\deltaI\subseteq V$ and
$\deltaO\subseteq V$. If $\Q$ is the I/O complexity for $C$ and $\Q'$ is
the I/O complexity for $C'$ then,
$\Q$ can be bounded by $\Q'$ as follows (tagging):
\begin{equation}
\label{eq:tag}
\Q'-|\deltaI|-|\deltaO|\leq \Q
\end{equation}
Reciprocally, $\Q'$ can be bounded by $\Q$ as follows (untagging):
\begin{equation}
\label{eq:untag}\Q \leq \Q'
\end{equation}
\end{theorem}
%\proofsk
%To prove the I/O tagging part, consider a valid game instance $\cal P$ for $C$, of cost $\IO(C)$. We
%can build a valid game instance $\cal P'$ for $C'$ from $\cal P$, with cost no more than
%$\IO(C)+|\deltaI|+|\deltaO|$, by replacing the transition $R3$ of any
%vertex $v\in \deltaI$ ($v\in \deltaO$) with transition $R1$ ($R2$) in $\cal P'$.
%I/O untagging can be proved in the similar fashion.
%\myendproof
%% \proof
%% Refer appendix~\ref{ap:io-compar}
%% \myendproof
%

\proof Consider an optimal calculation $\cal P$ for $C$, of cost $\Q$. We
will build a valid complete calculation $\cal P'$ for $C'$, of cost no more than
$\Q+|\deltaI|+|\deltaO|$. This will prove that $\Q' \leq
\Q+|\deltaI|+|\deltaO|$. We build $\cal P'$ from $\cal P$ as follows: (1)
for any input vertex $v\in \deltaI$, the (only) transition $R3$
involving $v$ in $\cal P$ is replaced in $\cal P'$ by a transition $R1$;
%(2 -- optional) for any input vertex $v\in I-I"$, any transition $R2$
%involving $v$ is not reported in $\cal P'$;
(2) for any output vertex $v\in \deltaO$,
% optional: such that at the end of $P'$ there is no blue pebble on it,
the (only) transition $R3$ involving $v$ in $\cal P$ is complemented
by an $R2$ transition;  (3) any other transition in $\cal P$ is reported as is in $\cal
P'$.

Consider now an optimal calculation $\cal P'$ for $C'$, of cost $\Q'$.
We will build a valid complete calculation $\cal P$ for $C$, of cost no
more than $\Q'$.
This will prove that $\Q\leq \Q'$. We build $\cal P$ from $\cal P'$ as
follows:
%(1) for any input vertex $v\in I-I"$ such that there is a single transition
%$R1$ involving $v$ in $\cal P'$, this transition is replaced by a transition
%$R3$;
(1) for any input vertex $v\in \deltaI$, the first transition $R1$ involving
$v$ in $\cal P'$ is replaced in $\cal P$ by a transition $R3$ followed by a
transition $R2$; (2) any other transition in $\cal P'$ is reported as is in
$\cal P$.
\myendproof

We note that such a construction is only possible for the restricted pebble
game where repebbling is disallowed. It enables tighter
lower bounds to be developed via CDAG decomposition. 
%For the R-MCL
%algorithm, each sub-CDAG corresponding to an iteration of the while
%loop can be associated with an I/O lower bound of
%$\Omega(N^3/\sqrt{S})$,
%to give a total lower bound of $\Omega(N^3T/\sqrt{S})$, where, $T$ is
%the number of iterations
%executed. 
In the next section, we use S-NR partitioning
and the untagging theorem in developing a static analysis approach 
to characterizing data-access lower bounds of loop programs.

%% file: static-analysis-lb.tex
\section{Parametric Lower Bounds via Static Analysis of Programs} 
\def\Dep{{D}\xspace}
\def\Smt{{V_F}\xspace}
\def\Flow{{F}\xspace}
\def\Son{S1\xspace}
\def\Stw{S2\xspace} 
\def\Sth{S3\xspace}
\def\Sfo{S4\xspace}
\def\U{U\xspace}
\def\V{V\xspace}
\def\C{C\xspace}
\def\R{\mathbb{R}}
\def\isl{\textsf{isl}}
\newcommand{\image}[1]{\mathsf{image}(#1)}
\newcommand{\domain}[1]{\mathsf{domain}(#1)}
\def\BIG{{\cal B}}
\def\round{\textsf{round}}
\def\true{\textsf{true}}
\def\false{\textsf{false}}
\newcommand\dimension[1]{\dim(#1)}
\newcommand\relation[1]{\mathsf{relation}(#1)}
\newcommand\computeSCC[1]{\textsf{compute\_SCC}(#1)}
\newcommand\scc[1]{\textsf{scc}(#1)}
\newcommand\ray[1]{\textsf{ray}(#1)}
\newcommand\try[1]{\textsf{try}(#1)}
\newcommand\broadcast[1]{\textsf{rkernel}(#1)}
\newcommand\base[1]{\textsf{base}(#1)}
\newcommand\solve[1]{\textsf{solve}(#1)}
\newcommand\best[1]{\textsf{best}(#1)}
\newcommand\vertices[1]{\textsf{vertices}(#1)}
\newcommand\cover[1]{\textsf{cover}(#1)}
\newcommand\simplify[1]{\textsf{simplify}(#1)}
\newcommand\subspace[1]{\textsf{subspace}(#1)}
\newcommand\objective[1]{\textsf{objective}(#1)}
\newcommand\constraint[1]{\textsf{constraint}(#1)}

\label{sec:static-analysis}
In this section, we develop a static analysis approach to derive 
asymptotic parametric I/O lower bounds as a function of cache size and problem
size, for \emph{affine computations}. Affine computations
can be modeled using (union of) convex sets of integer points, and
(union of) relations between these sets. The motivation is twofold.
First, there exists an important class of affine computations
whose control and data flow can be modeled exactly at compile-time using only
affine forms of the loop iterators surrounding the computation
statements, and program parameters (constants whose values are unknown
at compile-time). Many dense linear algebra computations, image
processing algorithms, finite difference methods, etc., belong to this
class of programs \cite{girbal.06.ijpp}.  Second, there exist readily
available tools to perform complex geometric operations on such sets and
relations. We use the Integer Set Library (ISL) \cite{verdoolaege2010isl}
for our analysis. 

In Subsection \ref{sec:sa-background}, we provide a description of the
program representation for affine programs.
In Subsection \ref{sec:sa-geom}, we detail the geometric reasoning that
is the basis for the developed I/O lower bounds approach. Subsection
\ref{sec:sa-automated} describes the I/O lower bound
computation using examples.

\subsection{\label{sec:sa-background}Background and Program Representation}

In the following, we use ISL terminology \cite{verdoolaege2010isl} and
syntax to describe sets and relations. We now recall some key concepts
to represent program features.

\paragraph{Iteration domain}
A computation vertex in a CDAG represents a dynamic instance
of some operation in the input program. For example,
given a statement $S1:$ \texttt{A[i] += B[i+1]} 
surrounded by one loop \texttt{for(i = 0; i < n; ++i)}, the operation
\texttt{+=} will be executed $n$ times, and each such dynamic
instance of the statement corresponds to a vertex in the CDAG.
For affine programs, this set of dynamic instances can be compactly
represented as a (union of) $\mathcal{Z}$-polyhedra, i.e., a set of
integer points bounded by affine inequalities intersected with
an affine integer lattice \cite{Gupta07zpoly}. Using ISL notation, the
iteration domain of statement $S1$, $D_{S1}$, is denoted:
\verb+[n]->{S1[i]:0<=i<n}+. The left-hand side of \verb+->+,
\verb+[n]+ in the example, is the list of all parameters needed to define the
set. \verb+S1[i]+ models a set with one dimension (\verb+[i]+) named
$i$, and the set space is named \verb+S1+. Presburger
formulae are used on the right-hand side of \verb+:+ to model the points
belonging to the set. In ISL, these sets are disjunctions of
conjunctions of Presburger formulae, thereby modeling unions of convex
and strided integer sets. The \emph{dimension} of a set $S$ is denoted as
$\dim(S)$. $\dim(S1) = 1$ in the example above. The \emph{cardinality} of set $S$ is denoted
as $|S|$. $|S1| = n$ for the example. Standard operations on sets,
such as union, intersection, projection along certain
dimensions, are available. In addition, key operations for analysis, such as
building counting polynomials for the set (i.e., polynomials of the
program parameters that model how many integer points are contained in
a set; $n$ in our example) \cite{barvinok.94.dcg}, and parametric
(integer) linear programming \cite{Fea88} are possible on such
sets. These operations are available in ISL.

We remark that although our analysis relies on integer
sets and their associated operations, it is not limited to programs
that can be exactly captured using such sets (e.g., purely affine
programs). Since we are interested in computing lower bounds on I/O,
an under-approximation of the statement domain and/or
the set of dependences is acceptable, since
an I/O  lower bound for the approximated system is a valid lower bound
for the actual system. For instance, if
the iteration domain $D_S$ of a statement $S$ is not described exactly
using Presburger formulae, we can under-approximate this set by
taking the largest convex polyhedron $\underline{D_S} \subseteq
D_S$. Such a polyhedron can be obtained, for instance, by first computing the
convex hull $\overline{D_S} \supseteq D_S$ 
%(then $D_R \subseteq \overline{D_R}$) 
and then
shifting its faces until they are strictly included in $D_S$. We also remark
that such sets can be extracted from an arbitrary CDAG (again using
approximations) by means of trace analysis, and especially trace
compression techniques for vertices modeling the same computation~\cite{ketterlin-micro12}.

%\paragraph{Connecting vertices}
\paragraph{Relations}
In the graph $G=(V,E)$ of a CDAG $C=(I,V,E,O)$, vertices are connected by producer-consumer edges capturing
the data flow between operations. Similar to iteration domains,
affine forms are used to model the \emph{relations} between
the points in two sets. Such
relations capture which data is accessed by a dynamic
instance of a statement, as in classical data-flow analysis. In the
example above, elements of array \texttt{B} are read in statement $S1$, and the relation $R1$
describing this access is:
\texttt{[n]->\string{S1[i] -> B[i+1] : 0<=i<n\string}}. 
This relation models a single edge between each element of set $S1$
and an element of set $B$, described by the relationship $i
\rightarrow i + 1$. Several operations on relations,
such as $\domain{R}$, which computes the domain (e.g., input set) of the
relation ($\domain{R1} =$ \texttt{[n] -> \string{[i] : 0<=i<n\string}}), $\image{R}$ computing
the image (e.g., range, or output set) of $R$ ($\image{R1} =$ \texttt{[n] ->
\string{[i] : 1<=i<n+1\string}}), the composition of two relations $R1 \circ R2$,
their union $\cup$, intersection $\cap$, difference $\setminus$ and
the transitive closure $R^+$ of a relation, are available. All these
operations are supported by ISL.

Relations can also be used to directly capture the connections between
computation vertices. For instance, given two statements $S1$ and $S2$
with a producer-consumer relationship, the
edges connecting each dynamic instance of $S1$ and $S2$ in a CDAG can be expressed using
relations. For example, 
\texttt{[n]->\string{S1[i,j] -> S2[i,j-1,k] : ...\string}} models a
relation between a 2D statement and a 3D statement. Each point in $S1$ is
connected to several points in $S2$ along the $k$-dimension.

We note that in a similar manner to iteration domains for vertices, relations
can also be extracted from non-affine programs via convex
under-approximation or from the CDAG via trace analysis. Again,
care must be taken to always properly under-approximate the relations
capturing data dependences: it is safe to ignore a dependence (it can
only lead to under-approximation of the data flow and therefore the I/O
requirement), and therefore we only consider must-dependences in
our analysis framework. 

\subsection{\label{sec:sa-geom}Geometric Reasoning for I/O Lower Bounds by 2S-partitioning}
Given a CDAG, Lemma ~\ref{lemma:hk.rbw} establishes a relation
between 
a lower bound on its data movement complexity for execution with $S$
fast storage elements and
the minimal possible number of vertex sets among all valid
$2S^{NR}$-partitions of the CDAG. The minimum possible number of
vertex sets in a $2S^{NR}$-partition is inversely related
to 
the largest possible size of any vertex set for a valid 
$2S^{NR}$-partition. A geometric reasoning based on the 
Loomis-Whitney inequality~\cite{lw49} and its generalization
\cite{HBL-ref,valdi} 
has been used to establish I/O lower bounds for
a number of linear algebra algorithms \cite{toledo.jpdc, BDHS11,Ballard:2012:BAS:2312005.2312021,Demmel2013TR}.
A novel approach to determining I/O lower bounds for affine computations
in perfectly nested loops has been recently developed \cite{Demmel2013TR}
using similar geometric reasoning. The approach developed in this paper
is inspired by that work and also uses a similar geometric reasoning, but improves on the prior work
in two significant ways:
\begin{compactenum}
\item {\em Generality}: It can be applied to a broader class of
computations, handling multiple statements and imperfectly nested loops.
\item {\em Tighter Bounds}: For computations with loop-carried
dependences that are not oriented perfectly along one of the iteration
space dimensions, it
provides tighter I/O lower bounds, as illustrated
by the Jacobi example in the next section.
\end{compactenum}
Before presenting the details of the static analysis 
for lower bounds characterization of affine computations,
we use a simple example to illustrate the geometric approach based
on the Loomis-Whitney inequality and its generalizations that have been
used to develop I/O lower bounds for matrix-multiplication and other
linear algebra computations. 
Consider the code exemplifying an N-body force calculation in 
Fig.~ \ref{fig:example-geom-reasoning}(a). We have a 2D iteration
space with $\Theta(N^2)$ points. %%%Each on $N$ particles
The net force
on each of $N$ particles from the other particles is computed
using the function \texttt{f()}, which uses the mass and position of a pair 
of particles to compute the force between them.
The total
number of input data elements for the computation is $\Theta(N)$. If
$\S < N$,
%the number of fast memory elements, 
it will be necessary to bring in
at least some of the input data elements more than once from slow to
fast memory.
\def\pp{{\textsf P\xspace}}
A geometric reasoning for a lower bound on the amount of I/O to/from
fast memory proceeds as follows. Consider an arbitrary vertex set from
any valid $2S^{NR}$-partition. Let the set of
points \pp\ in the iteration space, illustrated by a cloud in
Fig.~ \ref{fig:example-geom-reasoning}(b), denote the vertex set. The projections
of each of the points onto the two iteration space axes are shown.
Let $|\pp_i|$ and $|\pp_j|$ respectively denote the number of distinct
points on the $i$ and $j$ axes. $|\pp_i|$ represents the number
of distinct elements of input arrays \texttt{pos} and \texttt{mass} that are accessed
in the computation, for references \texttt{pos(i)} and \texttt{mass(i)}. Similarly,
$|\pp_j|$ corresponds to the number of distinct elements accessed
via the references \texttt{pos(j)} and \texttt{mass(j)}. For any vertex set from a valid 
$2S^{NR}$-partition, the size of the input set cannot exceed $2S$. Hence
$2 \times |\pp_i| \le 2S$ and $2 \times |\pp_j| \le 2S$.
For this 2D example, the Loomis-Whitney inequality asserts
that the number of points in \pp\ cannot exceed
$|\pp_i| \times |\pp_j|$. Combining the two inequalities, we
can conclude that $S^2$ is an upper bound on the size of the vertex set. Thus,
the minimum number of vertex sets in a valid 2S-partition, $H=\Omega(N^2/S^2)$.
By Lemma \ref{lemma:hk.rbw}, a lower bound on I/O is  $(H-1)\times \S$,
i.e., $\Omega(N^2/S)$.

\newsavebox{\fifthlisting}
\begin{lrbox}{\fifthlisting}
\lstset{language=C,basicstyle=\ttfamily}
{\small
\begin{lstlisting}
for(i=0;i<N;i++)
 for(j=0;j<N;j++)
  if (i <> j) force(i) 
  += f(mass(i),mass(j),pos(i),pos(j));
\end{lstlisting}
}
\end{lrbox}

\begin{figure}[h!tb]
\centering
\addtolength{\subfigcapskip}{1em}
\subfigure[Code for N-body force calculation] {\usebox{\fifthlisting}} 
\subfigure[Geometric Projection] {\raisebox{-1cm}{\includegraphics[width=3.0cm,viewport = 0 0 400 400]{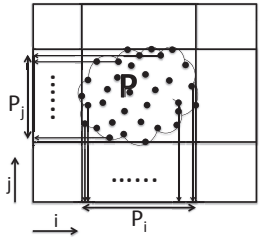}}}
\caption{\label{fig:example-geom-reasoning} Illustration of Geometric Reasoning for I/O Lower Bounds}
\end{figure}

More generally, for a $d$-dimensional iteration space, given some bounds
on the number of elements on some projections of $\pp$, a bound
on $|\pp|$ can be derived using a powerful approach developed by Christ et al.
~\cite{Demmel2013TR}. 
Christ et al. \cite[Theorem 3.2]{Demmel2013TR} extended the discrete
case of the Brascamp-Lieb inequality \cite[Theorem 2.4]{HBL-ref} to obtain
these bounds.
Since our goal here is to develop asymptotic parametric bounds, the
extension of the continuous Brascamp-Lieb inequality, stated below (in the
restricted case of orthogonal projections and using the Lebesgue
measure for volumes), is sufficient for our analysis.
We use the notation $H\le \R^d$ to denote that $H$ is a linear subspace of
$\R^d$.
\begin{theorem}
\label{thm.BL}
Let
$\phi_j:\ \R^d \rightarrow \R^{d_j}$ be an orthogonal projection for $j
\in \{1,2,\dots,m\}$ such that
$\phi_j(x_1,\dots,x_d)=(y_1,\dots,y_{d_j})$
where $\{y_1,\dots,y_{d_j}\}\subseteq \{x_1,\dots,x_d \}$.\\
Then, for $(s_1,\dots,s_m)\in [0,1]^m$:
\begin{eqnarray}
\forall H \le \R^d,\ \dim(H)\leq \sum_{j=1}^m s_j\dim(\phi_j(H))\hspace{3cm}\label{eq.const.inf}\\
\hspace{2cm}\Longleftrightarrow \hspace{1cm}\forall E\subseteq \R^d,\ |E|\leq \prod_{j=1}^m |\phi_j(E)|^{s_j}
\end{eqnarray}
\end{theorem}

Since the linear transformations $\phi_j$ are orthogonal projections, the
following Theorem enables us to limit the number of inequalities of
Eq.~(\ref{eq.const.inf}) required for Theorem~\ref{thm.BL} to hold. Only
one inequality per subspace $H_i$, defined as the linear span of the
canonical vector $e_i$, is required ($\left<e_i\right>$ represents
the subspace spanned by the vector with a non-zero only in the $i^{th}$
coordinate):
\begin{theorem}
\label{thm:product-case}
Let
$\phi_j:\ \R^d \rightarrow \R^{d_j}$ be an orthogonal projection for $j
\in \{1,2,\dots,m\}$ such that
$\phi_j(x_1,\dots,x_d)=(y_1,\dots,y_{d_j})$
where $\{y_1,\dots,y_{d_j}\}\subseteq \{x_1,\dots,x_d \}$.\\
Then, for $(s_1,\dots,s_m)\in [0,1]^m:$\\ 
\begin{eqnarray}
\forall H \le \R^d,\ \dim(H)\leq \sum_{j=1}^m s_j\dim(\phi_j(H))\hspace{3cm}\\
\hspace{-1cm}\Longleftrightarrow \hspace{1cm}
\forall H_i=\left<e_i\right>,\ 1=\dim(H_i)\leq \sum_{j=1}^m
s_i\delta_{i,j}\label{eq:const.finite}
\end{eqnarray}
where, $\delta_{i,j}=\dim(\phi_j(H_i))$
\end{theorem} 
The proof of Theorem \ref{thm:product-case} directly corresponds to the proof of
\cite[Theorem 6.6]{Demmel2013TR} and is omitted 
(see also \cite[Prop. 7.1]{HBL-ref}).
It shows that if $s=(s_1,\dots,s_m)\in [0,1]^m$ are such that 
$\forall H_i,\ 1\leq \sum_{j=1}^m s_i\delta_{i,j}$, then the volume of any measurable set 
$E\subseteq \R^d$ can be bounded by $U_s=\prod_{j=1}^m |\phi_j(E)|^{s_j}$.
In order to obtain as tight an asymptotic bound as possible, we seek $s$ such that
$U_s$ is as small as possible.
Since we have $|\phi_j(H)|\leq \S$, this corresponds to finding $s_j$ such that
$\prod_{j=1}^m S^{s_j}$ is minimized, or equivalently, $S^{\sum_{j=1}^m s_j}$ is minimized. 
In other words, $\sum_{j=1}^m s_j$ has to be minimized.
For this purpose, if $\forall i,\ \exists j,\ \textrm{s.t.,}\ \delta_{i,j}=1$, we solve:
\begin{equation}
\label{eq:min}
\text{Minimize }\sum_{j=1}^m s_j\text{,~~s.t.,~~}\forall i,\ 1\leq \sum_{j=1}^m
s_j\delta_{i,j}
\end{equation}
We can instead solve the following dual problem,
whose solution gives an indication of the shape of the optimal ``cube.''
\begin{equation}
\label{eq:max}
\text{Maximize }\sum_{i=1}^d x_i \text{,~~s.t.,~~}\forall j,\ \sum_{i=1}^d
x_i\delta_{i,j}\leq 1
\end{equation}
We use an illustrative example:
\lstset{language=C,basicstyle=\ttfamily}
{\small
\begin{lstlisting}
for(i=0;i<N;i++)
 for(j=0;j<N;j++)
  for(k=0;k<N;k++)
   C[i][j] = C[i][j] + A[i][k]*B[k]
\end{lstlisting}
}
Consider the following three projections
(we explain how the projection directions are obtained later
in this section)
\begin{equation*}
\phi_1: (i,j,k)\rightarrow (i,j);\
\phi_2:(i,j,k)\rightarrow (i,k);\
\phi_3:(i,j,k)\rightarrow (k)
\end{equation*}
Let $H_1$, $H_2$ and $H_3$ denote the three subspaces spanned by the
canonical bases of $\R^3$.
Consider, for example, the linear map $\phi_1$.
We have $\delta_{1,1}=\dim(\phi_1(h_1))=1$ for any $h_1\in H_1$,
$\delta_{2,1}=\dim(\phi_1(h_2))=1$ for any  $h_2\in H_2$, and
$\delta_{3,1}=\dim(\phi_1(h_3))=0$ for any $h_3 \in H_3$. Thus, we obtain the constraint
$x_1.1 + x_2.1 + x_3.0 \le 1$, or $x_1 + x_2 \le 1$. Similarly, we
obtain the remaining two constraints for the projections $\phi_2$ and
$\phi_3$.

This results in the following linear programming problem:
\begin{equation}
\label{eq:ex1}
\text{Maximize } x_1+x_2+x_3
\end{equation}
s.t. \hfill$x_1+x_2\leq 1;\ x_1+x_3\leq 1;\ x_3 \leq 1$\hfill~\\

Solving Eq. (\ref{eq:ex1}) provides the solution $(x_1,x_2,x_3)=(0,1,1)$,
i.e., $x_1+x_2+x_3=2$.
The solution corresponds to considering a cube of
asymptotic dimensions $1\times S \times S$ and volume $O(S^{\sum_{j=1}^3
x_j})$ =$O(S^2)$ as the largest vertex-set.
This provides an I/O lower bound of $\Omega(N^3/S)$, when the problem
size $N$ is sufficiently large.

\subsection{\label{sec:sa-automated}Automated I/O Lower Bound Computation}
We present a static analysis algorithm for automated derivation
of expressions for parametric asymptotic I/O lower bounds for programs. We
use two illustrative examples to explain the various steps in the algorithm
before providing detailed pseudo-code for the algorithm.
%
%We recall the following graph theory terminologies: a \emph{path} in a
%graph is a finite sequence of edges that connect distinct vertices;
%a \emph{circuit} is a finite sequence of distinct edges, that begins and
%ends at the same vertex.

\input{jacobi-1d}

\input{cooked}

\paragraph{Putting it all together:}
Algorithm~\ref{alg:node} provides a pseudo-code for our algorithm.
Because the number of possible paths in a graph is highly combinatorial,
several choices are made to limit the overall practical complexity
of the algorithm.  First, only edges of interest, i.e., those that correspond to relations
whose image is representative of the iteration domain, are kept.  Second,
paths are considered in the order of decreasing expected profitability.
One criterion detailed here corresponds to favoring injective circuits over 
broadcast paths with one-dimensional kernel (to reduce the potential
span),  and then broadcast paths with decreasing kernel dimension (the
higher the kernel, the more the
reuse, the lower the constraint). 

For a given vertex $v$, once the directions associated with the set of
paths chosen so far span the complete space of the domain of $v$, no
more paths are considered.
%Once, for a given
%vertex $A$, a set of path is selected, the amount of possible
%combinations is combinatorial in the dimension of the surrounding loop
%nest.  Though the combinatorial is reasonable, the number of considered
%path for this pass should be kept reasonable.  For this purpose lots of
%paths are dropped off at this step.  In particular if the dimension
%along which the reuse appears is already represented by another path
%(which can be checked quite easily without any polyhedron manipulations)
%we can give up with this one.  
The role of the function \try{} (on lines 20, 26 and 32 in
Algorithm~\ref{alg:node}) amounts to finding a set of paths that are
linearly independent, compatible (i.e., a base can be associated to
them), and representative.  The funtion \try{} is shown in
Algorithm~\ref{alg:try}. The function \best{$v$} (shown in
Algorithm~\ref{alg:best})
selects a set of paths for a vertex $v$ and computes the associated
complexity. The function \solve{} (shown in Algorithm~\ref{alg:solve})
writes the linear program and returns the I/O lower bound (with cases)
for a domain $D$ and a set of compatible subspaces.

%For a given vertex $A$,
%to a set of paths associated to $A$ corresponds a sub-graph and an \IO.
%% The goal of \cover{} is to solve the weighted cover problem of $G_F$
%by sub-graphs weighted by their \IO.  % It is performed greedily
%starting from sub-graphs for which the order of magnitude of the \IO
%(only asymptotic formulas are manipulated) are largest first, then the
%one with the smallest number of vertices, and then using a topological
%order of the control flow graph.
%In more details, for the pseudo-code we suppose that:
\noindent Various operations used in the pseudo-code are detailed below.
\begin{compactitem}
\item Given a relation $R$, $\domain{R}$ and $\image{R}$ return the
domain and image of $R$, respectively.

\item For an edge $e$, the operation $\relation{e}$ provides its
associated relation. If $R_e=\relation{e}$
has acceptable number of disjunctions, then the edge can be split into
multiple edges with count equal to the number of disjunctions,
otherwise, a convex under-approximation can be done.

\item For a given path $p=(e_1,e_2,\dots,e_l)$ with associated relations
$(R_{e_1},R_{e_2},\dots,R_{e_l})$ we can compute the associated
relation for $p$ by composing the relations of its edges, i.e.,
$\relation{p}$ computes
$R_{e_l}\circ\dots \circ R_{e_2}\circ R_{e_1}$.
Note that the domain of the composition of two
relations is restricted to the points for which the composition can
apply, i.e. $\domain{R_i\circ
R_j}=R_j^{-1}(\image{R_j}\cap\domain{R_i})$ and $\image{R_i\circ
R_j}=R_i(\image{R_j}\cap\domain{R_i})$.

\item For a given domain $D$, $\dim(D)$ returns its dimension.
If the cardinality of $D$ (i.e., number of points
in $D$) is represented in terms of the program parameters, its
dimension can be obtained  by setting the values of the
parameters to a fixed big value (say $\BIG$), and computing
$\log_B(\card{D})$, and rounding the result to the nearest integer. For
example, if $|D|=C(n,m)=nm+n+3$, setting $\BIG=10^3$, we get
$\dim(D)=\round\left(\log_{\BIG}\left(C(\BIG, \BIG)\right)\right)=2$.

\item If a relation $R$ is injective and can be expressed as an affine
map of the form $\mat{A}.\vctr{x}+\vctr{b}$, then the operation
$\ray{R}$ computes $\vctr{b}$, otherwise, returns $\bot$.

\item For a relation $R$, if its inverse can be expressed as an affine
relation $\mat{A}.\vctr{x}+\vctr{b}$, $\broadcast{R}$ computes the
kernel of the matrix $A$
(and returns $\bot$ otherwise).

\item For a set of vectors $b=\{\vec{b_1},\dots, \vec{b_l}\}$,
$\subspace{b}$ provides the linear subspace spanned by those vectors.

\item For a set of linear subspaces $K=\{k_1,\dots,k_l\}$, $\base{K}$
gives a set of linearly independent vectors
$b=\{\vec{b_1},\dots,\vec{b_d}\}$ such that for any $k_i$, there exists
$b_i\subseteq b$ s.t. $k_i=\subspace{b_i}$. If such a set could not be
computed, it returns $\bot$.

\item For a path $p$, $\vertices{p}$ returns its set of vertices.

\item Given an expression $X$, the operation $\simplify{X}$ simplifies
the expression by eliminating the lower order terms. For example,
$\simplify{NT + N^2 -N + T}$ returns $NT+N^2$.
\end{compactitem}

\begin{algorithm}[h!tb] 
\small
$G_F=(V_F,E_F)$\;
$F_I:=\emptyset$;~~$F_B:=\emptyset$;~~$F_{BB}:=\emptyset$\;
\ForEach{$e=(u,v)\in E_F$}{
  $R :=\relation{e}$\;
  \lIf{$\dimension{\image{R}}<\dimension{v}$}{\textbf{next}}
  \lIf{$R$ is invertible}{$F_I:=F\cup \{e\}$}
  \If{$\dimension{\domain{R}}=\dimension{\image{R}}-1$}{$F_B:=F_B\cup \{e\}$}
  \If{$\dimension{\domain{R}}<\dimension{\image{R}}-1$}{$F_{BB}:=F_{BB}\cup \{e\}$}
}
%% $\computeSCC{F_I}$\\

\ForEach{$v\in \Smt$}{
  $d:=\dimension{v}$\;
  \ForEach{circuit $p$ from $v$ to $v$ in $F_I$}{
    $R:=\relation{p}$\;
    \lIf{$(b:=\ray{R})=\bot$}{\textbf{next} $p$}
    \lIf{$\dimension{\image{R}}<d$}{\textbf{next} $p$}
    \lIf{$\try{v, \subspace{b}, p}$}{\textbf{next} $v$}
  }
  \ForEach{cycle-free path $p$ to $v$ in $F_{B}F_I^*$}{
    $R:=\relation{p}$\;
    \lIf{$(k:=\broadcast{R})=\bot$}{\textbf{next} $p$}
    \lIf{$\dimension{\image{R}}<d$}{\textbf{next} $p$}
    \lIf{$\try{v, k, p}$}{\textbf{next} $v$}
  }      
  \ForEach{cycle-free path $p$ to $v$ in $F_{BB}F_I^*$}{
    $R:=\relation{p}$\;
    \lIf{$(k:=\broadcast{R})=\bot$}{\textbf{next} $p$}
    \lIf{$\dimension{\image{R}<d}$}{\textbf{next} $p$}
    \lIf{$\try{v, k, p}$}{\textbf{next} $v$}
  }
  $\best{v}$\;
}
$\simplify{\sum_{v\in\Smt} v.\textit{complexity}}$\;
\caption{\label{alg:node}For each vertex $v$ in a data-flow graph $G_F$, finds a set of paths and computes the corresponding complexity.}
\end{algorithm}

\begin{algorithm}[h!tb] 
\small
\textbf{Function} \best{vertex $v$}\\
\textsf{let} $(k,K,D,T)\in\textit{v.clique}$ with maximum lexicographic value of $\left(\dim(D), \textrm{dimension}(k), -\sum_{k_i\in K} \textrm{dimension}(k_i), \solve{D,K}, -|T|\right)$\;
$Q:=\solve{D, K}$\;
$v.\textit{complexity}:=Q$\;
\caption{\label{alg:best}For a vertex $v$, selects a set of paths and computes the
associated complexity.}
\end{algorithm}

\begin{algorithm}[h!tb] 
\small
\textbf{Function} \try{vertex $v$, subspace $k'$, path $p$}\\
\{\textit{v.clique} is a set of quadruples $(k,K,D,T)$ where:
\begin{tabular}{lll}
&- & $k$ is a subspace, \\
&- &$K$ is a set of subspaces, \\
&- & $D$ is a domain, \\
&- & $T$ is a set of vertices\}
\end{tabular}\\
\{\textit{v.complexity} is an asymptotic complexity (with cases)\}\\
\lIf{$k'\in \textit{v.subspaces}$}{\textbf{return} \false}
$\textit{v.subspaces} := \textit{v.subspaces} \cup \{k'\}$\;
\ForEach{$(k,K,D,T)\in \textit{v.clique}\cup (\bot,\bot,\bot,\bot)$}{
  \If{$\textrm{dimension}(k+k')>\textrm{dimension}(k)$ \textbf{and}
    ~~~~~~~~~~~~~~~~~~$\base{K\cup\{k'\}}\neq\bot$}{
    $D'=\image{\relation{p}}\cap D$\;
    \If{$D=\bot$ \textbf{or} $\dimension{D'}=\dimension{D}$}{
      $T':=T\cup \vertices{p}$\;
      $K':=K\cup\{k'\}$\;
      $\textit{v.clique}:=\textit{v.clique}\bigcup(k+k',K',D',T')$\;
      \If{$\textrm{dimension}(k+k')\geq \dimension{\domain{v}}$}{
        $Q:=\solve{D', K'}$\;
        $v.\textit{complexity}:=Q$\;
        \textbf{return} \true;
      }
    }
  } 
}
\textbf{return} \false;

\caption{\label{alg:try}For a vertex $v$, try to add path $p$ to some other paths. Return true if a good bound is found.}
\end{algorithm}

\begin{algorithm}[h!tb]
\small
\textbf{Function} \solve{domain $D$, set of subspaces $K$}\\
$b:=\base{K}$\;
$LP:=\objective{\textrm{maximize} \Theta=\sum_{b_i\in b} \alpha_i}$\; 
\lForEach{$k\in K$}{ $LP:=LP . \constraint{\sum_{b_i\not\in k}\alpha_i\leq 1}$}
\ForEach{$b'\subset b$}{
  $D_{b'}:=\textsf{projection}(\subspace{b'}, D)$\;
  $LP:=LP . \constraint{\sum_{b_i\in b'}\alpha_i\leq \log_S(|D_{b'}|)}$\;
}
$F:=\sum_{k\in K}|\textsf{projection}(\neg k,D)|$\;
$\Theta:=\textsf{solution}(LP)$\;
$U:=S^\Theta$\;
$Q:=\Omega\left(\frac{|D|S}{U}\right)-\Omega(F)$\;
\textbf{return} $Q$\;
\caption{For a domain $D$ and a set of compatible subspaces, writes the
linear program, and returns the I/O lower bound (with cases).\label{alg:solve}}
\end{algorithm}

%% \begin{algorithm}[h]
%% \small
%% \textbf{Function} \cover{set of statements $S$}\\
%% $Q:=O(|\textrm{Inputs}|)|+O(|\textrm{Outputs})|)$\;
%% %% \ForEach{$A\in S$ sorted with lexicographic value of $(Q_A, -|T_A|, topological\_CFG(A))$ with $(Q_A,T_A)=A.\textit{Cover}$}{
  %% \If{$T_A\subset S$}{
    %% $Q:=(Q,Q_A)$\;
    %% $S:=S\backslash T_A$\;
  %% }
%% }
%% \textbf{return} $Q$\;
%% %% \caption{Given a set of sub-graphs and their corresponding \IO, cover $S$ with disjoint sub-graphs trying to maximizing the overall \IO}   
%% \end{algorithm}

%% file: jacobi-1d.tex
\paragraph{Illustrative example 1:}
\newcommand\ceil[1]{\left\lceil#1\right\rceil}
\newcommand{\dom}[1]{D_{#1}}
\def\Z{\mathcal{Z}\xspace}
\def\I{{E}\xspace}
Consider the following example of Jacobi 1D stencil computation.
\lstset{language=C,basicstyle=\ttfamily}
{\small
\begin{lstlisting}
Parameters: N, T
Inputs: I[N]
Outputs: A[N]
   for (i=0; i<N; i++)
S1:  A[i] = I[i];

   for (t=1; t<T; t++)
   {
     for (i=1; i<N-1; i++)
S2:    B[i] = A[i-1] + A[i] + A[i+1];
    
     for (i=1; i<N-1; i++)
S3:    A[i] = B[i];
   }
\end{lstlisting}
}

\begin{figure}[h!tb]
\centering
\includegraphics[width=4cm]{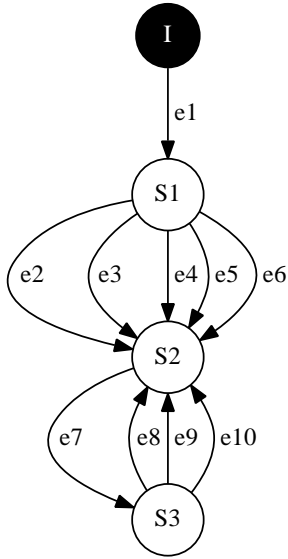}
\caption{\label{fig:jacobi1d-gf}Data-flow graph for Jacobi 1D}
\end{figure}

Fig. \ref{fig:jacobi1d-gf} shows the static data-flow graph
$G_F=(V_F,E_F)$ for Jacobi 1D. $G_F$ contains a vertex for
each statement in the code. The input array \verb+I+ is also explicitly
represented in $G_F$ by node $I$ (shaded in black 
in Fig.
\ref{fig:jacobi1d-gf}).
Each vertex has an associated domain as shown below:
\begin{compactitem}
\item $D_{I}=$\verb+[N]->{I[i]:0<=i<N}+
\item $D_{S1}=$\verb+[N]->{S1[i]:0<=i<N}+
\item $D_{S2}=$\verb+[T,N]->{S2[t,i]:1<=t<T and 1<=i<N-1}+
\item $D_{S3}=$\verb+[T,N]->{S3[t,i]:1<=t<T and 1<=i<N-1}+
\end{compactitem}
The edges represent the true (read-after-write) data dependences
between the statements. Each edge has an associated affine dependence
relation as shown below:
\begin{compactitem}
\item Edge $e1$: This edge corresponds to the dependence due to copying the inputs \verb|I| to
	array \verb|A| at statement $S1$ and has the following
	relation.\\
	\verb+[N]->{I[i]->S1[i]:0<=i<N}+
\item Edges $e2$, $e3$ and $e4$: The use of array
	elements \verb+A[i-1], A[i]+ and \verb|A[i+1]| at statement
	$S2$ are captured by edges $e2$, $e3$ and $e4$, respectively.
\begin{verbatim}
[T,N]->{S1[i]->S2[1,i+1]:1<=i<N-2}
[T,N]->{S1[i]->S2[1,i]:1<=i<N-1}
[T,N]->{S1[i]->S2[1,i-1]:2<=i<N-1}
\end{verbatim}
\item Edges $e5$ and $e6$: Multiple uses of the boundary elements \verb+I[0]+
	and \verb+I[N-1]+ by \verb+A[t][1]+ and \verb+A[t][N-2]+,
	respectively, for
	\verb+1<=t<T+ are represented by the following relations.\\
	\verb|[T,N]->{S1[0]->S2[t,1]:1<=t<T}|\\
	\verb|[T,N]->{S1[N-1]->S2[t,N-2]:1<=t<T}|
\item Edge $e7$: The use of array \verb+B+ in statement \verb+S3+
	corresponds to edge $e7$ with the following relation.\\
	\verb|[T,N]->{S2[t,i]->S3[t,i]:1<=t<T and 1<=i<N-1}|
\item Edges $e8$, $e9$ and $e10$: The uses of array \verb+A+ in statement
	$S2$ from $S3$ are represented by these edges with the following
	relations.
\begin{verbatim}
[T,N]->{S3[t,i]->S2[t+1,i+1]:1<=t<T-1 and 1<=i<N-2}
[T,N]->{S3[t,i]->S2[t+1,i]:1<=t<T-1 and 1<=i<N-1}
[T,N]->{S3[t,i]->S2[t+1,i-1]:1<=t<T-1 and 2<=i<N-1}
\end{verbatim}
%\item Edge e11: Values of array \verb+A+ being used later as output is
%	captured through an edge from vertex $S3$ to $A$ and has the following
%	associated relation.\\
%	\verb|[T,N]->{S3[T-1,i]->A[i]:1<=i<N-1}|
\end{compactitem}
Given a path $p=(e_1,\dots, e_l)$ with associated edge relations
$(R_1, \dots,R_l)$, the relation associated with
$p$ can be computed by composing the relations of its edges, i.e.,
$\relation{p}=R_l\circ \cdots \circ R_1$. 
For instance, the relation for the path $(e7,e8)$ in the example, obtained through
the composition $R_{e8}\circ R_{e7}$, is given by
$R_p=$ \texttt{[T,N] -> \string{S2[t,i] -> S2[t+1,i+1]\string}}.
Further, the domain and image of a
composition are restricted to the points for which
the composition can apply, i.e., $\domain{R_j\circ
R_i}={R_i}^{-1}(\image{R_i}\cap \domain{R_j})$ and $\image{R_j\circ
R_i}=R_j(\image{R_i}\cap \domain{R_j})$. Hence, $\domain{R_p}=$
\texttt{[T,N] -> \string{S2[t,i] : 1<=t<T-1 and 1<=i<N-2\string}} and
$\image{R_p}=$ \texttt{[T,N] -> \string{S2[t,i] : 2<=t<T and
2<=i<N-1\string}}.

Two kinds of paths, namely, \emph{injective circuit} and \emph{broadcast path},
defined below, are of specific importance to the analysis.
\begin{definition}[Injective edge and circuit]
An \emph{injective edge} $a$ is an edge of a data-flow graph whose
associated relation $R_{a}$ is both affine and injective, i.e.,
$R_{a}=\mat{A}.\vctr{x}+\vctr{b}$, where $\mat{A}$ is an invertible matrix.
An \emph{injective circuit} is a circuit $E$ of a data-flow graph such
that every edge $e\in E$ is an injective edge.
\end{definition}
\begin{definition}[Broadcast edge and path]
A \emph{broadcast edge} $b$ is an edge of a data-flow graph whose associated
relation $R_{b}$ is affine and $\dim(\domain{R_{b}}) <
\dim(\image{R_{b}})$. A \emph{broadcast path} is a path
$(e_1,\dots,e_n)$ of a data-flow graph such that $e_1$ is a broadcast
edge and $\forall_{i=2}^{n} e_i$ are injective edges.
\end{definition}
Injective circuits and broadcast paths in a data-flow graph
essentially indicate multiple uses of same data, and therefore are
good candidates for lower bound analysis. Hence only 
paths of these two kinds are considered in the analysis. The current example of
Jacobi 1D computation illustrates the use of injective circuits to
derive I/O lower bounds, while the use of broadcast paths for
lower bound analysis is explained in another example that follows.
\noindent\textbf{Injective circuits:}
In the Jacobi example, we have three circuits
%$p1=(e7,e8),~p2=(e7,e9)$ and $p3=(e7,e10)$ 
to vertex $S2$ through $S3$. The relation for each circuit
is computed by composing the relations of its edges as explained
earlier. The relations, and the dependence vectors they represent, are
listed below.
\begin{compactitem}
\item Circuit $c_1=(e7,e8)$:\\
	$R_{c_1}=$ \texttt{[T,N] -> \string{S2[t,i]->S2[t+1,i+1] : 1<=t<T-1
	and 1<=i<N-2\string}}\\
	$\vctr{b_1}=(1,1)^T$
\item Circuit $c_2=(e7,e9)$:\\
	$R_{c_2}=$ \texttt{[T,N] -> \string{S2[t,i]->S2[t+1,i] : 1<=t<T-1 and
	1<=i<N-1\string}}\\
	$\vctr{b_2}=(1,0)^T$
\item Circuit $c_3=(e7,e10)$:\\
	$R_{c_3}=$ \texttt{[T,N] -> \string{S2[t,i]->S2[t+1,i-1] : 1<=t<T-1
	and 2<=i<N-1\string}}\\
	$\vctr{b_3}=(1,-1)^T$
\end{compactitem}
%Corresponding to each of these three circuits, $c_1,~c_2$ and $c_3$,
%in $G_F$, there is a set of paths in $G$, say, $P_1,~P_2$ and $P_3$,
%respectively. 
%
\begin{figure}
\begin{center}
	\includegraphics[width=\columnwidth]{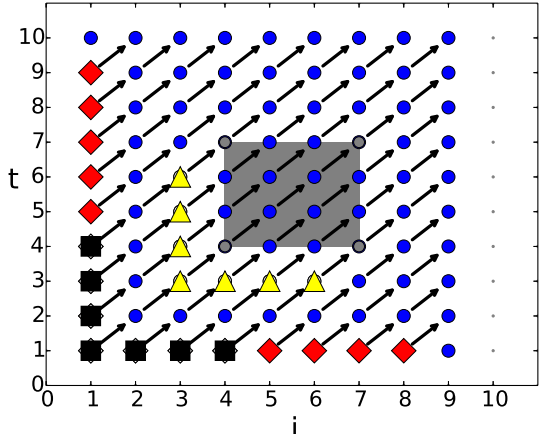}
\caption{\label{fig:jacobi1d-iterdom}Original iteration domain space for
Jacobi 1D. Blue circles: Integer points of domain $D_{S2}$; Black
arrows: Relation $R_{c1}$ of circuit $(e7,e8)$; Red diamonds: Frontier
$F_1$; Gray box: subset $\I$ and corresponding vertex-set $\upsilon_1$;
Yellow triangles:
$\In{\upsilon_1}$; Black squares: Projection of the points inside gray box
along the direction of black arrows onto the frontier.}
\end{center}
\end{figure}
%\begin{figure}
%\begin{center}
%\begin{pspicture}(5,5)
%\psline(0,0)(5,5)
%\end{pspicture}
%\caption{\label{fig:jacobi1d-transformed}Transformed space}
%\end{center}
%\end{figure}
%Given a vertex $v\in V_F$ and a list of circuits that begin and end at
%$v$, we restrict the domain of $v$ to the intersection of domains of its
%circuit relations. Hence, here we have $D_{S2}=\domain{R_{c_1}} \cap
%\domain{R_{c_2}} \cap\domain{R_{c_3}}=$ \texttt{[T,N] -> \{S2[t,i] : 
%1<=t<T-1 and 2<=i<N-2\}}.
Fig. \ref{fig:jacobi1d-iterdom} pictorially
shows the domain $D_{S2}$ and the relation $R_{c_1}$
as a $\Z$-polyhedron for $T=N=11$. 
%We could also think of this
%$\Z$-polyhedron as the sub-CDAG $\C_1=(I_1,V_1,E_1,O_1)$ corresponding to the sub-graph
%$G_{F_1}=(\{S2,S3\}, \{e7,e8,e9,e10\})$ being embedded into the
%iteration domain with each integer point representing a ``macro-node''
%consisting of -- an instance of $S2$, $S3$ and the edge between them.
%Hence, each edge in the $\Z$-polyhedron coincides with an edge from an
%instance of $S3$ to $S2$ in $G_1=(V_1,E_1)$.

\begin{definition}[Frontier]
The frontier, $F$, of a relation $R$, with domain $D$, is the set of
points with no incoming edges in the corresponding $\Z$-polyhedron.
%representing $D$ and $R$. 
\end{definition}
$F$ can be calculated using the set operation $F = D \setminus R(D)$.
The frontiers $F_1$, $F_2$ and $F_3$ for the three relations, $R_{c_1}$, $R_{c_2}$ and
$R_{c_3}$ respectively, are listed below.
\begin{compactitem}
\item $F_1=$ \verb|[T,N]->{S2[1,i]:2<=i<N-2; S2[t,1]:1<=t<T-1}|
\item $F_2=$ \verb|[T,N]->{S2[1,i]:1<=i<N-1}|
\item $F_3=$ \verb|[T,N]->{S2[1,i]:2<=i<N-1; S2[t,N-2]:2<=t<T-1}|
\end{compactitem}
In Fig. \ref{fig:jacobi1d-iterdom}, points of frontier $F_1$ are shown
as red diamond shaped points.
Due to the correspondence between a $\Z$-polyhedron and a (sub-)CDAG (refer
to Sec.~\ref{sec:sa-background}), each
point in a frontier represents a source vertex (i.e., vertex with no
incoming edges) of the (sub-)CDAG. It could be seen that there are
$\card{F_i}, 1\le i \le 3$, disjoint paths $P_i,1\le i \le 3$ (as a consequence of the
injective property of the relations) in the sub-CDAG
$C_1=(I_1,V_1,E_1,O_1)$
(corresponding to the instances of statements $S2$ and $S3$), each with a distinct
source vertex that corresponds to a point in $F_i$. These
source vertices are tagged as inputs for the lower bounds analysis and
their count, $\card{F_i}$, is later
subtracted from the final I/O lower bound using Theorem
\ref{thm:io-compar}.
%Furthermore,
%projecting the set of integer points representing the vertices in the
%$\Z$-polyhedron along the direction of an edge leads us to the 
%set of points that forms the frontier. In Fig.
%\ref{fig:jacobi1d-iterdom} the points corresponding to the frontier
%$F_1$, obtained by projecting the integer points along the direction
%$\vctr{b_1}=(1,1)$ (red colored edges in
%\ref{fig:jacobi1d-iterdom}), are shown with red circles around them.

%Now, consider a complete calculation $\P_1$ of $\C_1$. 
Let $\upsilon_1\subseteq V_1$ be a
vertex-set of a valid $2\S$-partition of $\C_1$. There are a set of points
$\I$ in the $\Z$-polyhedron (e.g., the set of points inside the gray
colored box in Fig.
\ref{fig:jacobi1d-iterdom}) corresponding to $\upsilon_1$. The set of points
outside $\I$ with an edge to a point in $\I$ corresponds to
$\In{\upsilon_1}$ (e.g., points marked with yellow triangles in Fig.
\ref{fig:jacobi1d-iterdom}). Since there is no cyclic dependence between the vertex-sets
of the 2\S-partition and the paths are disjoint, by starting from 
the vertices of $\In{\upsilon_1}$ and tracing backwards along the paths
in any $P_i,~1\le i \le 3$, we should reach
$\card{\In{\upsilon_1}}\le 2\S$ distinct source vertices.
This process 
corresponds to projecting the set $\I$ along each of the directions
$\vctr{b_i},~1\le i \le 3$ onto the frontier $F_i,~1\le i \le 3$.  
Hence
we have $\card{\I_{\downarrow \vctr{b_i}}}\le 2\S$ (here,
$\I_{\downarrow \vctr{b_i}}$ denotes projecting $\I$ along the
direction $\vctr{b_i}$). 
The
points of the frontier obtained by projection are shown as black
squares (over red diamonds) in Fig.~\ref{fig:jacobi1d-iterdom}.
We ensure that $\I \subseteq D^R_{S2}=\domain{R_{c_1}} \cap
\image{R_{c_1}}\cap \domain{R_{c_2}} \cap\image{R_{c_2}}
\cap\domain{R_{c_3}}\cap\image{R_{c_3}}$.
This allows us to apply the geometric
reasoning discussed in Sec. \ref{sec:sa-geom} to restrict the size of the
set $\I$ as shown below.
%by constraining the size of the projections $\card{\I_{\downarrow
%\vctr{b_i}}}$.
%as shown in Thm. \ref{thm:injection-lb}.
Since $\dim(D_{S2})=2$, it is
sufficient to consider any two linearly independent directions.

Theorem \ref{thm:product-case}
applies only for
projections along the orthogonal directions.
In case projection vectors are non-orthogonal, a simple change of basis
operation is used to transform the space to a new space where the
projection directions are the canonical bases. In the example, if we
consider vectors $\vctr{b_1}/\card{\vctr{b_1}}$ and
$\vctr{b_3}/\card{\vctr{b_3}}$ as the projection
directions in the original space, then the linear map
$\left(\begin{array}{cc}
%		1 & 1\\
%		1 & -1\\
		\vctr{b_1}/\card{\vctr{b_1}} & \vctr{b_3}/\card{\vctr{b_3}}\\
\end{array}\right)^{-1}$ 
will transform the $\Z$-polyhedron to a new
space where the projection directions are the canonical bases. 
%Then, the following theorem allows us to obtain the I/O lower bound on
%$\C_1$.
%%Fig.  \ref{fig:jacobi1d-transformed} shows the transformed space. 
%%
%%In the following theorem, the $\Z'$-polyhedron represents such a
%%polyhedron in the transformed space.
%\begin{theorem}[Injective circuits and I/O lower bound]
%	\label{thm:injection-lb}
%	Let $C=(I,V,E,O)$ be a CDAG.
%	Let $D$ be the domain 
%	of the $\Z$-polyhedron corresponding to
%	$C$. Let $\{\vctr{b_1},\dots,\vctr{b_m}\}$ be set of orthogonal
%	projection directions obtained from the injective circuits (after
%	performing the change-of-basis operation, if needed).
%	If the linear constraints
%	(\ref{eq:const.finite}) of Thm. \ref{thm:product-case} are
%	feasible, then for sufficiently large $\card{D}$, the minimum I/O
%	required to compute $C$ is
%	$\Omega\left(\frac{\card{D}}{\S^{\Theta-1}}\right)$, where, $\Theta=\min \sum_{j=1}^m
%s_j$ and $\S$ is the number of red pebbles.
%\end{theorem}
%\proof
%Let $(V_1,V_2,\dots,V_h)$ be a valid 2$\S^{NR}$-partition of $C$.
%Let $\I\subseteq D$ be a set of points corresponding to the vertices in
%any $V_i,1\le i \le h$. Then, we have $\card{\I}\le \prod_{j=1}^m
%|\phi_j(E)|^{s_j}$ for any $s=(s_1,\dots,s_m)$ satisfying
%(\ref{eq:const.finite}).
%We also have that $\card{\phi_j(E)} \le 2\S$ as detailed above.
%\myendproof
%
In the example, after such transformation, the projection vectors
are $(1,0)^T$ and $(0,1)^T$, and hence we have the following two
projections: $\phi_1:(i,j) \rightarrow (i)$; $\phi_2:(i,j) \rightarrow
(j)$.
From Eq. \ref{eq:const.finite}, we obtain the following inequalities for the
dual problem (refer (\ref{eq:max})): $x_1\le 1$; $x_2\le 1$. 

In
addition, we also need to include constraints for the degenerate
cases where the problem size considered may be small relative to the
cache size, $\S$. Hence, we have the following additional constraints for the
example:
$\card{\phi_1(\I)}^{x_1}\le (N+T)$; $\card{\phi_2(\I)}^{x_2}\le
(N+T)$, or (after taking $\log$ with base $\S$), $x_1
\log_S(\card{\phi_1(E)}) \le
\log_S(N+T)$; $x_2 \log_S(\card{\phi_1(E)})) \le \log_S(N+T)$. Since $\card{\phi_j(E)}\le
\S$, we have $\log_S(\card{\phi_j(E)})\le 1$. 
%This constant
%can be ignored for our asymptotic bounds. 
Hence, we obtain the
constraints $x_1\le \log_S(N+T)$ and $x_2 \le \log_S(N+T)$. Thus, we
solve the following following parametric linear programming problem.
%
%Finally, by including constraints to handle the degenerate cases
%$\left(\card{\I_{\downarrow \vctr{b_1}}}\le N+T,~\card{\I_{\downarrow
%\vctr{b_3}}}\le N+T\right)$, we end
%up with the following parametric linear programming problem.
\begin{equation}\label{eq:jacobi1d}
	\text{Maximize }\Theta=x_1 + x_2
\end{equation}
\begin{eqnarray*}
\text{s.t.~~~~~~~~}x_1 & \le & 1\\
	x_2 & \le & 1\\
	x_1 & \le & \log_S(N+T)\\
	x_2 & \le & \log_S(N+T)\\
\end{eqnarray*}
Solving Eq. (\ref{eq:jacobi1d}) using PIP \cite{Fea88} provides the
following solution:\\
If $\log_S(N+T) \ge 1$ then, $ x_1 = x_2 = 1 $, else, $
x_1 = x_2 = \log_S(N+T)$. 

This specifies that when $N+T =
\Omega(\S)$, $\card{\upsilon_1}=O(S^2)$, and hence
$\Q=\Omega\left(\frac{NT}{S}-(N+T)\right)$ (here, $(N+T)$ is
subtracted from the lower bound to account for I/O tagging), otherwise,
$\card{\upsilon_1}=O\left((N+T)^2\right)$ and
$\Q=\Omega\left(\frac{NTS}{(N+T)^2}-(N+T)\right)$.
%~\par
%\noindent{\textbf{Broadcast paths:}}
%The edges $e5$ and $e6$ of the data-flow graph correspond to broadcast
%edges. We are specifically interested in broadcast paths whose reverse
%of their relations can be expressed as affine relations. For instance,
%the edge $e_5$ has the associated relation $R_{e_5}=$\texttt{[T,N] ->
%\string{S1[i] -> S2[t,i+1] : 1<=t<T and i=0\string}}. $R_{e_5}^{-1}$ can be
%expressed as the following affine function:
%$$
%\left(\begin{array}{cc}0 & 1\end{array}\right)
%.\left(\begin{array}{c}t\\ i\end{array}\right)
%+\left(\begin{array}{c}0\\ -1\end{array}\right)
%$$
%The dimension of the kernel of the matrix $A=\left(\begin{array}{cc}0 & 1\end{array}\right)$ is one. In general, the
%dimension of the kernel should be greater than or equal to one. As in
%the case of injective circuit, we tag the frontier,
%$F_5=R_{e_5}^{-1}(\domain{R_{e_5}})$ as inputs
%and use the
%basis vectors of the kernel as the projection directions.

In the
example, since the vectors $\vctr{b_1}/\card{\vctr{b_1}}$ and
$\vctr{b_3}/\card{\vctr{b_3}}$ are already
orthonormal, the change of basis transformation that we performed
earlier is unimodular. But, in general this
need not be the case. 
%This transformation need not be always unimodular.
Since we focus only on 
asymptotic parametric bounds, 
%we can just use the original set of linearly independent
%directions for projection and 
any constant multiplicative factors
that arise due to the non-unimodular transformation are ignored.

%% file: cooked.tex
\paragraph{Illustrative example 2:}
The following example is composed of a scaled matrix-multiplication and
a Gauss-Seidel computation within an outer iteration loop.
\lstset{language=C,basicstyle=\ttfamily}
{\small
\begin{lstlisting}
Parameters: W, N, T
Inputs: A[N][N], C[N][N], Temp[N][N]
Outputs: A[N][N], C[N][N]
   // Iterative loop with scaled Matmult
   // followed by Stencil
   for(it=0;it<W;it++)
   {
     // Scaled Matmult split out into a sequence of 
     // mat-vec and vector scaling ops for each row
     for(i=0;i<N;i++)
     {
       for(j=0;j<N;j++)
         for(k=0;k<N;k++)
S1:        Temp[i][j] += A[i][k]*A[k][j];
   
       for(j=0;j<N;j++)
S2:      Temp[i][j] = 2*Temp[i][j];
   
       for(j=0;j<N;j++)
S3:      C[i][j] += Temp[i][j];
     }
   
     // Seidel stencil
     for(t=0;t<T;t++)
       for (i=1; i<N-1; i++)
         for (j=1; j<N-1; j++)
S4:        A[i][j] = 0.5 * (A[i-1][j] + A[i][j-1] \
             + A[i][j] + A[i+1][j] + A[i][j+1]);
   }
\end{lstlisting}
}

The decomposition theorem (Theorem \ref{thm:decomposition}) allows us to split
this code into individual components, analyze each sub-program
separately and obtain the I/O lower bounds for the whole program through
simple summation of the individual bounds. Hence, given the CDAG $C$ of
the above example, the analysis proceeds with the following steps:
\begin{compactitem}
\item The CDAG $C$ and thus the underlying program is decomposed as
follows: (1) Each iteration of the outer loop, with trip-count
\texttt{W}, is split into $W$ sub-programs. (2) Each of this
sub-program is further decomposed by separating the matmult (consisting
of statements $S1,~S2$ and $S3$) and Seidel
operations (consisting of statement $S4$) into individual sub-programs.
\item The vertices corresponding to the input arrays of the matmult and
Seidel computations are tagged as inputs in their corresponding sub-CDAGs.
\item The matmult (with sub-CDAG
$C_{m}=(I_{m}, V_{m}, E_{m}, O_{m})$) and the Seidel computation
(with sub-CDAG $C_{s}=(I_{s}, V_{s}, E_{s}, O_{s})$) are separately
analyzed for their I/O lower bounds.
\item If $L_{m}$ and $L_{s}$ are the I/O lower bounds obtained in the
previous step for matmult and Seidel computation, respectively,
Theorem
\ref{thm:decomposition} and Theorem \ref{thm:io-compar} provides us an I/O
lower bound of\\
$\Omega\left(W\times \left((L_{m}-\card{I_{m}})+(L_{s}-\card{I_{s}})\right)\right)$
for the whole program.
\end{compactitem}
The analysis of the Seidel computation is similar to the analysis
of the Jacobi 1D computation detailed in the previous example. Hence, we
skip the analysis and provide the following result: If $N =
\Omega(\sqrt{S})$ and $T=\Omega(\sqrt{S})$ then,
$\Q_s=\Omega\left(\frac{N^2T}{\sqrt{S}}-N^2-NT\right)$, else
%$\Q_s=\Omega\left(\frac{NTS}{(N+T)^2}-(N+T)\right)$, 
$\Q_s\ge 0$.
where, $Q_s$ is
the I/O complexity for the Seidel computation.

Now, we consider the analysis of the scaled matmult.
The data-flow graph, $G_F$ consists of six vertices: vertices $A$,
$C$ and
$Temp$ correspond to the input arrays \texttt{A}, \texttt{C} and
\texttt{Temp}, respectively; vertices
$S1$, $S2$ and $S3$ correspond to the statements $S1$, $S2$ and $S3$,
respectively.
The domain corresponding to each vertex (in the order $A,~C,~Temp,~S1,~S2$ and
$S3$) is listed below:
\begin{compactitem}
\item $D_A=$\texttt{[N]->\string{A[i,j]:0<=i<N and 0<=j<N\string}}
%\item $D_B=$\texttt{[N]->\string{B[i,j]:0<=i<N and 0<=j<N\string}}
\item $D_C=$\texttt{[N]->\string{C[i,j]:0<=i<N and 0<=j<N\string}}
\item $D_{Temp}=$\texttt{[N]->\string{Temp[i,j]:0<=i<N and 0<=j<N\string}}
\item $D_{S1}=$\texttt{[N]->\string{S1[i,j,k]:0<=i<N and 0<=j<N and 0<=k<N\string}}
\item $D_{S2}=$\texttt{[N]->\string{S2[i,j]:0<=i<N and 0<=j<N\string}}
\item $D_{S3}=$\texttt{[N]->\string{S3[i,j]:0<=i<N and 0<=j<N\string}}
\end{compactitem}
The relations corresponding to various edges are listed below.
\begin{compactitem}
\item $\relation{e1=(A,S1)}=R_{e1}=$
\texttt{[N] -> \string{A[i, j] -> \\S1[i,j',j] : 0<=i<N and 0<=j<N and 0<=j'<N\string}}
\item $\relation{e2=(A,S1)}=R_{e2}=$
\texttt{[N] -> \string{A[i,j] -> \\S1[i',j,i] : 0<=i'<N and 0<=i<N and 0<=j<N\string}}
\item $\relation{e3=(C,S3)}=R_{e3}=$
\texttt{[N] -> \string{C[i,j] -> S3[i,j] : 0<=i<N and 0<=j<N\string}}
\item $\relation{e4=(Temp,S1)}=R_{e4}=$
\texttt{[N] -> \string{Temp[i,j] -> S1[i,j,0] : 0<=i<N and 0<=j<N\string}}
\item $\relation{e5=(S1,S1)}=R_{e5}=$
\texttt{[N] -> \string{S1[i,j,k] -> \\S1[i,j,k+1] : 0<=i<N and 0<=j<N and 0<=k<N-1\string}}
\item $\relation{e6=(S1,S2)}=R_{e6}=$
\texttt{[N] -> \string{S1[i,j,N-1] -> S2[i,j] : 0<=i<N and 0<=j<N\string}}
\item $\relation{e7=(S2,S3)}=R_{e7}=$
\texttt{[N] -> \string{S2[i,j] -> S3[i,j] : 0<=i<N and 0<=j<N\string}}
\end{compactitem}
\textbf{Broadcast paths:} The paths $p_1=(e1)$ and $p_2=(e2)$
are of type broadcast. As $p_1$ and $p_2$ are
composed of a single edge, their relations, $R_{p_1}$ and $R_{p_2}$
respectively, are the same as their edge. Thus, $R_{p_1}=R_{e1}$ and
$R_{p_2}=R_{e2}$.
%On the other hand, 
%the relation for path $p_3$, $R_{p_3}$, is a composition of $R_{e4}$ and
%the transitive closure over $R_{e5}$; since the edge $e5$ is a
%circuit, transitive closure of $R_{e5}$ is required to capture the chain
%of edges in the corresponding sub-CDAG. Thus, $R_{p_1}=R_{e1}$;
%$R_{p_2}=R_{e2}$; $R_{p_3}=R_{e5}^+\circ R_{e4}=$
%\texttt{[N] -> \string{Temp[i,j] -> S1[i,j,l] : 0 <=i<N and 0<=j<N and
%1<=l<N\string}}.
We are specifically interested in the broadcast paths whose
inverse-relations (e.g., $R_{p_1}^{-1}$)
can be expressed as affine maps. In our example, the two inverse-relations
$R_{p_1}^{-1}$ and $R_{p_2}^{-1}$ can be expressed as affine maps as
shown below:
\begin{eqnarray*}
%R_{p_1}^{-1}: &\left(
%\begin{array}{c}
%i \\ j
%\end{array}
%\right)=
R_{p_1}^{-1} & \equiv &
\left(
\begin{array}{ccc}
1 & 0 & 0\\
0 & 0 & 1\\
\end{array}
\right).
\left(
\begin{array}{c}
i \\ j' \\ j
\end{array}
\right)+
\left(
\begin{array}{c}
0 \\ 0 \\ 0
\end{array}
\right)
\\
%R_{p_2}^{-1}: &\left(
%\begin{array}{c}
%i \\ j
%\end{array}
%\right)=
R_{p_2}^{-1} & \equiv &
\left(
\begin{array}{ccc}
0 & 0 & 1\\
0 & 1 & 0\\
\end{array}
\right).
\left(
\begin{array}{c}
i' \\ j \\ i
\end{array}
\right)+
\left(
\begin{array}{c}
0 \\ 0 \\ 0
\end{array}
\right)
\\
%R_{p_3}^{-1}: &\left(
%\begin{array}{c}
%i \\ j
%\end{array}
%\right)=
%R_{p_3}^{-1} & \equiv &
%\left(
%\begin{array}{ccc}
%1 & 0 & 0\\
%0 & 1 & 0\\
%\end{array}
%\right).
%\left(
%\begin{array}{c}
%i \\ j \\ l
%\end{array}
%\right)+
%\left(
%\begin{array}{c}
%0 \\ 0 \\ 0
%\end{array}
%\right)
\end{eqnarray*}
Further, we have an injective circuit $p_3=(e5)$ with $R_{p_3}=R_{e5}$, whose direction
vector $\vctr{b_3}=(0,0,1)^T$.

We next calculate the frontiers $F_1,~F_2$ and $F_3$ of the relations
$R_{p_1}$, $R_{p_2}$ and $R_{p_3}$, respectively, by taking the 
set-difference of their domain and image (e.g.,
$F_1=D_{p_1} \setminus R_{p_1}(D_{p_1})$), where,
$D_{p_1}=\domain{p_1}$).
%as we did for the injective
%circuit case.
The three frontiers are shown using the ISL notation below:
\begin{compactitem}
\item $F_1=$\texttt{[N] -> \string{A[i,j] : 0<=i<N and 0<=j<N\string}}
\item $F_2=$\texttt{[N] -> \string{A[i,j] : 0<=i<N and 0<=j<N\string}}
\item $F_3=$\texttt{[N] -> \string{S1[i,j,0] : 0<=i<N and 0<=j<N\string}}
\end{compactitem}

In the case an injective circuit (with associated relation, say, $R_a$),
we chose the direction of projection to be the vector representing $R_a$.
Here, in case of a broadcast path (with associated relation, say,
$R_b=\mat{A}.\vctr{x}+\vctr{b}$), we choose the kernel of the matrix
$\mat{A}$, $\ker(\mat{A})$, to be the projection direction. The intuition behind choosing
this direction is that the kernel represents the plane of reuse, and
hence, the set of points obtained by projecting a set $\I$ along the
kernel directions represents the $\In{\I}$.
%lies orthogonal to the subspace spanning the instances of source
%vertices. 
In general, the kernel can be of dimension higher than one
(but has to be at least one due to the definition of a broadcast path).
The kernels ($\vctr{k_1}$ and $\vctr{k_2}$) of the
inverse-relations of the paths $p_1$ and $p_2$
are: $\vctr{k_1}=(0,1,0)^T$ and $\vctr{k_2}=(1,0,0)^T$, respectively.
%As the kernel direction is along the reuse direction, the set of
%points obtained through projection of a set $\I$ along a kernel vector
%represents the $\In{\I}$.
%It is possible to show, by the same reasoning as in the case of
%injective circuits, that constraining the size of the projection of a
%set of points, say $\I$, along a kernel direction is
%equivalent to restricting the size of $\In{\I}$.
%As the set of projection directions obtained in our example are equal to
%the set of canonical bases, Thm. \ref{thm:product-case} is directly
%applicable. 

By choosing $\vctr{k_1}$, $\vctr{k_2}$ and $\vctr{b_3}$ as the
projection directions, we obtain $\phi_1:(i,j,k) \rightarrow (i,k)$;
$\phi_2:(i,j,k) \rightarrow (j,k)$; $\phi_3:(i,j,k) \rightarrow (i,j)$.
This provides us the following inequalities: $x_1+x_3 \le 1$; $x_2+x_3
\le 1$; $x_1+x_2 \le 1$. Further, to handle the degenerated cases, we
have the additional constraints that specify that the size of the
projections onto the subspaces $\{\vctr{i}\}$, $\{\vctr{j\}}$ and
\{$\vctr{k\}}$ should not exceed $N$ and the size of the projections
onto the subspaces $\{\vctr{i},\vctr{j}\}$, $\{\vctr{j},\vctr{k}\}$ and
$\{\vctr{i},\vctr{k}\}$ should not exceed $N^2$.
Hence, we obtain the following parametric linear
programming problem.
\begin{equation}
\label{eq:cooked}
\text{Maximize } \Theta = x_1 + x_2 + x_3
\end{equation}
\begin{eqnarray*}
\text{s.t.~~~~~~~~}x_1 + x_2 & \le & 1\\
x_2 + x_3 & \le & 1\\
x_1 + x_3 & \le & 1\\
x_1 & \le & \log_S(N)\\
x_2 & \le & \log_S(N)\\
x_3 & \le & \log_S(N)\\
x_1 + x_2 & \le & 2\log_S(N)\\
x_2 + x_3 & \le & 2\log_S(N)\\
x_1 + x_3 & \le & 2\log_S(N)\\
\end{eqnarray*}
Solving Eq. (\ref{eq:cooked}) using PIP \cite{Fea88} provides the
following solution:\\
If $2\log_S(N)\ge 1$ then, $x_1=x_2 \ =x_3 = 1/2$, else,
$x_1=x_2 \ =x_3 = \log_S(N)$.
Hence, when $N=\Omega(\sqrt{\S})$, $\Q_m =
\Omega\left(\frac{N^3}{\sqrt{\S}}-N^2\right)$, otherwise, $\Q_m\ge 0$.

Finally, by applying Theorem \ref{thm:decomposition}, we obtain the I/O
lower bound for the full program, $\Q \ge \Q_m + \Q_s =
\Omega\left(W\times\left(\frac{N^3}{\sqrt{\S}} + \frac{N^2T}{\sqrt{\S}} - N^2 -
NT\right)\right)$
when $N$ and $T$ are sufficiently large.

%% file: related.tex
%\vspace*{-3ex}
\section{Related Work}
\label{sec:related}
%\vspace*{-2ex}

\hk provided the first formalization of the I/O complexity problem
for a two-level memory hierarchy
using the red/blue pebble game on a CDAG and the equivalence to 2S-partitions
of the CDAG.
We perform an
adaptation of \hk 2S-partitioning to constrain the size of
the input set of each vertex set rather than a dominator set, which
is suitable for bounding the minimum I/O for a CDAG with the
restricted red/blue pebble game where repebbling is disallowed. This
adaptation enables effective composition of lower bounds of
sub-CDAGs to form I/O lower bounds for composite CDAGs. A similar
adaptation has previously been used by modifying
the red/blue pebble game through addition of a third kind of pebble \cite{elango-spaa2014,elango-TR2013}. The composition of lower bounds for sequences
of linear algebra operations
has previously been addressed by the work of
Ballard et al. \cite{BDHS11} 
by use of ``imposed'' reads and writes in between segments of operations, adding the lower bounds
on data access for each of the segments, and subtracting the number of imposed reads and writes.
Our use of tagged inputs and outputs in conjunction with application of the decomposition theorem
bears similarities to the use of imposed reads and writes by Ballard et al., but is applicable to the more
general model of CDAGs that model data dependences among operations.

\comment{
Bilardi et al. \cite{bilardi-tcs1999} developed a techniqe called
``closed-dichotomy size technique'' to obtain lower bounds on the
access complexity of a DAG in terms of space lower bounds; a key
aspect of that work is that the technique can be separately applied to
each member of a family of disjoint subsets of vertices of a DAG, when
recomputation is not allowed. Using the notion of free-input space
complexity, this approach was later extended to the
case when recomputation is allowed \cite{bilardi-gtccs2000}.
}

In \cite{bilardi-tcs1999} an approach is proposed to obtain e lower
bound to the access complexity of a DAG in terms of space lower bounds
that apply to disjoint components of the DAG, when recomputation is
not allowed.  In \cite{bilardi-gtccs2000}, the approach is extended to
the case when recomputation is allowed, by means of the notion of
free-input space complexity.

%Several others followed \hk's work on I/O complexity in deriving lower
%bounds on data accesses
%\cite{aggarwal.ca.88,AACS87,toledo.jpdc,bilardi2000,bilardi2001characterization,savage.cc.95,savage.book,ranjan11.fft,ranjan12.rpyr,valiant.jcss.11,DemmelGHL12,BDHS11,BDHS11a,Demmel2013TR,apsp_ipdps2013,savage.options}.
%Aggarwal et al. provided several lower bounds for sorting algorithms
%\cite{aggarwal.ca.88}.  Savage \cite{savage.cc.95,savage.book}
%developed an alternate notion of $S$-span of a graph and it has been
%used in a number of works to develop lower
%bounds \cite{ranjan11.fft,ranjan12.rpyr,savage.options}.
%{\bf \large \color{red} Add more lower bound papers here, including Toledo}
Irony et al.~\cite{toledo.jpdc} used a geometric reasoning with the
Loomis-Whitney inequality \cite{lw49} to present an alternate proof 
to Hong and Kung's \cite{hong.81.stoc} for I/O lower
bounds on standard matrix multiplication.
More recently, Demmel's group at UC Berkeley has developed lower bounds as well as
optimal algorithms for several linear algebra computations including
QR and LU decomposition and the all-pairs shortest paths problem
\cite{BDHS11,BDHS11a,DemmelGHL12,apsp_ipdps2013}.

Extending the scope of the \hk model to more complex memory
hierarchies has also been the subject of research. Savage provided an
extension together with results for some classes of computations that
were considered by \hk, providing optimal lower bounds for I/O with
memory hierarchies \cite{savage.cc.95}. Valiant proposed a
hierarchical computational model \cite{valiant.jcss.11} that offers
the possibility to reason in an arbitrarily complex parametrized
memory hierarchy model.  While we use a single-level memory model in
this paper, the work can be extended in a straight forward manner to
model multi-level memory hierarchies.

Unlike \hk's original model, several models have been proposed that do
not allow recomputation of values (also referred to as ``no
repebbling'')
\cite{BDHS11,BDHS11a,bilardi2001characterization,bilardi12-bsp,michele13,ranjan11.fft,savage.cc.95,savage.book,savage.options,cook.pebbling,toledo.jpdc,ranjan12.rpyr,ranjan12.vertex}.
Savage \cite{savage.cc.95} developed results for FFT using no
repebbling.  Bilardi and Peserico \cite{bilardi2001characterization}
explore the possibility of coding a given algorithm so that it is
efficiently portable across machines with different hierarchical
memory systems, without the use of recomputation.  Ballard et
al.~\cite{BDHS11,BDHS11a} assume no recomputation in deriving lower
bounds for linear algebra computations.  Ranjan et
al.~\cite{ranjan11.fft} develop better bounds than \hk for FFT using a
specialized technique adapted for FFT-style computations on memory
hierarchies. Ranjan et al.~\cite{ranjan12.rpyr} derive lower bounds
for pebbling r-pyramids under the assumption that there is no
recomputation. As discussed earlier, we also use a model 
that disallows recomputation of values. But our focus in this
regard is different
from previous efforts -- we formalize an adaptation of the
the 2S-partitioning model of \hk that facilitates effective
composition of lower bounds from sub-CDAGs of a composite CDAG.

\newsavebox{\thirdlisting}
\begin{lrbox}{\thirdlisting}
\lstset{language=C}
{\small
\begin{lstlisting}
for(i=0;i<N;i++)
 for(j=0;j<N;j++)
  for(k=0;k<N;k++)
   C[i][j]+=A[i][k]+B[k][j];
\end{lstlisting}
}
\end{lrbox}

\newsavebox{\fourthlisting}
\begin{lrbox}{\fourthlisting}
\lstset{language=C}
{\small
\begin{lstlisting}
for(i=0;i<N;i++)
 for(j=0;j<N;j++)
  for(k=0;k<N;k++)
  {
   C[i][j] += 1;
   A[i][k] += 1;
   B[k][j] += 1;
  }
\end{lstlisting}
}
\end{lrbox}

\begin{figure}[h!tb]
\vspace{-.5cm}
\centering
\addtolength{\subfigcapskip}{1em}
\subfigure[Matrix Multiplication Code] {\usebox{\thirdlisting}} 
\subfigure[Code with same array accesses as Mat-Mult] {\usebox{\fourthlisting}} 
\caption{\label{fig:Demmel-ex} Example illustrating difference between CDAG model and computational model used by Christ et al. \cite{Demmel2013TR}}
\end{figure}

The previously described efforts on I/O lower bounds have involved
manual analysis of algorithms to derive the bounds. In contrast,
in this paper we develop an approach to automate the analysis of
I/O lower bounds for programs. The only other such effort to our
knowledge is the recent work of Christ et al.  \cite{Demmel2013TR}.
Indeed, the approach we have develop in this paper was inspired 
by their work,
but differs in a number of significant ways:
\begin{compactenum}
\item The models of computation are different. Our work is based on the
CDAG and pebbling formalism of \hk, while the lower bound results
of Christ et al. \cite{Demmel2013TR} are based on a different 
abstraction of an indivisible loop body of
affine statements within a perfectly nested loop. For example,
under that model, the lower bounds for codes in Fig.~\ref{fig:Demmel-ex}(a)
(standard matrix multiplication) and Fig.~\ref{fig:Demmel-ex}(b)
would be exactly the same -- $O(N^3/\sqrt{S})$ -- since the analysis
is based only on the array accesses in the computation. 
In contrast, with the
red/blue pebble-game model, the CDAGs for the two codes are very
different, with the matrix-multiplication code in
Fig.~\ref{fig:Demmel-ex}(a) representing a connected CDAG,
while the code in Fig.~\ref{fig:Demmel-ex}(b) represents has a
CDAG with three disconnected parts corresponding to the three statements,
and computation has a much lower I/O complexity of $O(N^2)$.

\item The work of Christ et al.  \cite{Demmel2013TR} does not model data dependencies between
statement instances, and can therefore produce weak lower bounds.
In contrast, the approach developed in this paper is based on using precise data dependence
information as the basis for geometric reasoning in the iteration
space to derive the I/O lower bounds.
For example, with the 2D-Jacobi example discussed earlier, the lower
bound obtained by the approach of Christ et al. would be $O(N^2)$ instead of the tight
bound of $O(N^2T/\sqrt{S})$ that is obtained with the algorithm developed
in this paper. 

\item This work addresses a more general model of programs. While 
the work of Christ et al. \cite{Demmel2013TR} only models perfectly nested loop
computations, the algorithms presented in this paper handle
sequences of imperfectly nested loop computations. 
\end{compactenum}

%% file: disc.tex
%\vspace*{-3ex}
%\vspace{-.75cm}
\section{Discussion}
\label{sec:discussion}
%% {\large \color{red} Saday: I have made some edits to previous version: To Be Discussed}

%\vspace*{-2.5ex}
We conclude by raising some issues and open questions, some of which
are being addressed in ongoing work. %\\

\noindent{\bf Tightness of lower bounds}: A very important question is
whether a lower bound is tight -- clearly, zero is a valid but weak
and useless I/O lower bound for any CDAG. The primary means of
assessing tightness of lower bounds is by comparison with upper bounds
from algorithm implementations that have been optimized for data
locality.  For example, tiling (or blocking) is a commonly used
approach to enhance data locality of nested loop
computations. %Previous work has shown how to minimize the reuse
%distance between operations (i.e., \cite{uday08pldi}), but is limited
%to using affine schedules for the CDAG vertices, thereby not
%guaranteeing to find the I/O optimal schedule for the program.
%% FIXME: LNP: cannot say that anymore, we published the UB paper...
%% An open question is whether any
%% automated analysis for upper bounds can be developed to compare with
%% the generated lower bounds? 
An open question is whether any automatic tool can be designed to
systematically explore the space of valid schedules to
generate good parametric upper bounds based on models and/or heuristics
that minimize data movement cost.

%% {\bf \color{red} Need to review work on parametric data
%% locality optimization; \color{blue} Ram: any thoughts on this?}.\\

\noindent{\bf Lower bounds when recomputation is allowed}: The vast
majority of existing application codes do not perform any redundant
recomputation of any operations. But with data movement costs becoming
increasingly dominant over operation execution costs, both in terms of
energy and performance, there is significant interest in devising
implementations of algorithms where redundant recomputation of values
may be used to trade off additional inexpensive operations for a reduction
in expensive data movements to/from off-chip memory. It is therefore
of interest to develop automated techniques for I/O lower bounds under
the original model of \hk that permits re-computation of CDAG
vertices. Having lower bounds under both models can offer a mechanism
to identify which algorithms have a potential for a trade-off between
extra computations for reduced data movement and which do not.

If the CDAG representing a computation has matching and tight I/O
lower bounds under both the general model and the restricted model,
the algorithm does not have
potential for such a trade-off. On the other hand, if a lower bound
under the restricted model (that prohibits re-pebbling) is higher than
a tight lower bound under the general model, the computation has
potential for trading off extra computations for a reduction in volume
of data movement. 
%Interestingly, all published I/O lower bounds 
%for practically used algorithms are of the former kind, i.e., offering 
%no potential for benefit
%from re-computation. The only cases of algorithms where the lower
%bounds under the two models differ, is with composite computations
%comprised of steps with different kernels.
This raises an interesting question:
{\em Is it possible to develop necessary and/or sufficient conditions 
on properties of the computation, for example on the nature of
the data dependencies, which will guarantee
  matching (or differing) lower bounds under the
  models allowing/prohibiting re-computation?}

\noindent{\bf Relating I/O lower bounds to machine parameters}:
I/O lower bounds can be used to determine whether an algorithm will be
inherently limited to performance far below a processor's peak because
of data movement bottlenecks. The collective bandwidth between main
memory and the last level cache in multicore processors in words/second on
current systems is over an order of magnitude lower than
the aggregate computational performance of the processor cores in
floating-point operations per second; this ratio is a critical machine
balance parameter.  By comparing this machine balance parameter to the
ratio of the I/O lower bound (calculated for $S$ set to the capacity
of the last level on-chip cache) to the total number of arithmetic operations
in the computation, we can determine if the algorithm will be
inherently limited by data movement overheads. However, such an analysis
will also require tight assessment of the constants for the leading
terms in the asymptotic expressions of the order
complexity for I/O lower bounds. This is not addressed by
the approach presented in this paper.
%%{\bf \color{red} Fab: any thoughts on this?}

\noindent{\bf Modeling associative operators}: Reductions using associative
operators like addition occur frequently in computations. With the CDAG
model, some fixed order of execution is enforced for such computations,
resulting in an over-constrained linear chain of dependencies between the 
vertices corresponding to
instances of an associative operator. Some previously developed
geometric approaches to modeling I/O lower bounds 
\cite{toledo.jpdc,BDHS11,Demmel2013TR}
have developed I/O lower bounds for a family of algorithms that differ
in the order of execution of associative operations. It would be of interest
to extend the automated lower bounding approach of this paper to also model
lower bounds among a family of CDAGs corresponding to associative 
reordering of the operations.

\noindent{\bf Finding good decompositions}: The second illustrative example in Sec.~\ref{sec:static-analysis} demonstrated the benefit of judiciously
decomposing CDAGs to obtain good lower bounds by combining bounds
for sub-CDAGs via the decomposition theorem. However, if the decomposition is
performed poorly, the result will be a very weak lower bound. In the same 
example, if the computation within the second level \texttt{i} loop had also been used to further decompose
the CDAG, we would have a sequence of matrix-vector multiplications with
order complexity $O(N^2)$ from which the tagged I/O nodes of the same order
of complexity must be subtrated out, resulting in a weak lower bound of zero.
Conversely, if the computation within the outer \texttt{it} loop were not decomposed
into sub-CDAGs, it would again have resulted in weak lower bounds.
The question of automatically finding effective decompositions of CDAGs to
enable tight lower bounds is an open problem.

%% file: conclusion.tex
%\vspace*{-3ex}
%\vspace{-.5cm}
\section{Conclusion}
\label{sec:conclusion}
%\vspace{-.4cm}
%\vspace*{-2ex}
Characterizing the I/O complexity of a program is a cornerstone
problem, that is particularly important with current and emerging
power-constrained architectures where data movement costs are
the dominant energy bottleneck. Previous approaches to modeling the
I/O complexity of computations have several limitations that this paper has
addressed. First, by suitably modifying the pebble game model used for
characterizing I/O complexity, analysis of large composite
computational DAGs is enabled by decomposition into smaller sub-DAGs, a key
requirement to allow the analysis of complex programs. Second, 
a static analysis approach has been developed to compute I/O
lower bounds, by generating asymptotic parametric data-access lower bounds for
programs as a function of cache size and problem size.

%% alternate lower bounding technique to \hk
%% 2S-partitioning, using convex min-cut graph partitioning. Using the
%% developed techniques, the first I/O lower bounds analysis for the
%% recently developed CA-CG method was performed.
%\vspace{-.4cm}
%
%section moved to the main latex file
%\section*{Acknowledgment}